\documentclass{article}
\usepackage{ijcai26}

\usepackage{times}
\usepackage{soul}
\usepackage{url}
\usepackage[hidelinks]{hyperref}
\usepackage[utf8]{inputenc}
\usepackage[small]{caption}
\usepackage{graphicx}
\usepackage{amsmath}
\usepackage{amsfonts}
\usepackage{amssymb}
\usepackage{amsthm}
\usepackage{booktabs}
\usepackage{multirow}
\usepackage[switch]{lineno}
\usepackage[table]{xcolor}

\usepackage{xcolor}
\usepackage{algpseudocode}
\usepackage[ruled,vlined,linesnumbered,noend]{algorithm2e}

\newtheorem{theorem}{Theorem}
\newtheorem{assumption}{Assumption}

\title{Sparsification Under Siege: Dual-Level Defense Against Poisoning in Communication-Efficient Federated Learning}

\author{
Jin, Zhiyong$^1$
\and
Runhua Xu$^1$\and
Chao Li$^{3,4}$\and
Yizhong Liu$^2$\and
Jianxin Li$^{1,5}$\and
James Joshi$^6$\\
\affiliations
$^1$School of Computer Science and Engineering, Beihang University\\
$^2$School of Cyber Science and Technology at Beihang University\\
$^3$School of Cyberspace Science and Technology, Beijing Jiaotong University\\
$^4$Beijing Key Laboratory of Security and Privacy in Intelligent Transportation\\
$^5$Zhongguancun Laboratory\\
$^6$School of Computing and Information,University of Pittsburgh\\
\emails
\{sy2406106, runhua\}@buaa.edu.cn,
li.chao@bjtu.edu.cn,
liuyizhong@buaa.edu.cn,
lijx@buaa.edu.cn,
jjoshi@pitt.edu
}

\begin{document}

\maketitle

\begin{abstract}

Gradient sparsification, while mitigating communication bottlenecks in Federated Learning (FL), fundamentally alters the geometric landscape of model updates. 
We reveal that the resultant high-dimensional orthogonality renders traditional Euclidean-based robust aggregation metrics mathematically ambiguous, creating a ``sparsity-robustness trade-off'' that adversaries exploit to bypass detection. 
To resolve this structural dissonance, we propose SafeSparse, a consensus restoration framework that decouples defense into topological and semantic dimensions. 
Unlike prior arts that treat sparsification and security orthogonally, \textit{SafeSparse} introduces: 
(1) a Structure-Aware Calibration mechanism utilizing Jaccard similarity to filter topological outliers induced by index poisoning; 
and (2) a Directional Semantic Alignment module employing density-based clustering on update signs to neutralize magnitude-invariant attacks.
Theoretically, we establish convergence guarantees for \textit{SafeSparse}. 
Extensive experiments across multiple datasets and attack scenarios demonstrate that \textit{SafeSparse}  recovers up to 25.7\% global accuracy under coordinated poisoning, effectively closing the vulnerability gap in communication-efficient FL.
\end{abstract}

\section{Introduction}

Federated Learning (FL) has emerged as a paradigm shift in distributed artificial intelligence, enabling collaborative training across decentralized edge devices while preserving data locality~\cite{kairouz2021advances}. 
Despite its privacy benefits, the communication bottleneck inherent in transmitting high-dimensional parameter updates remains a critical hurdle. 
To address this, gradient sparsification techniques, such as Top-$k$ selection, have become the de facto standard, reducing uplink communication overheads by orders of magnitude (often exceeding 99\%) without significantly compromising convergence~\cite{sattler2019sparse,sattler2019robust,han2020adaptive}.

However, we argue that this pursuit of efficiency has inadvertently introduced a fundamental structural vulnerability. 
Existing robust aggregation protocols, such as Krum~\cite{blanchard2017machine}, Geometric Median (RFA)~\cite{pillutla2022robust}, and FLGuardian~\cite{zhou2025flguardian}, are predicated on the geometric assumption of \textit{Dense Euclidean Consensus}.
They assume that benign updates cluster around a global mean in a dense vector space, allowing outliers to be detected via Euclidean metrics ($L_2$ norm).

We reveal that Top-$k$ sparsification violates this premise.
Mathematically, the sparsification operator acts as a nonlinear projection that maps updates into low-dimensional subspaces. 
In non-IID settings, this results in benign updates becoming \textit{sparse orthogonal}, namely, their non-zero indices (masks) barely overlap. 
Consequently, Euclidean distance becomes mathematically ambiguous as a similarity metric; two benign clients holding disjoint but valid features may appear infinitely distant from each other. 
We term this phenomenon the \textbf{Sparsity-Robustness Trade-off}: the very mechanism used to save bandwidth destroys the geometric landscape required for security.
By manipulating the sparse index masks (Structure) rather than just the parameter values (Semantics), malicious clients can artificially concentrate their contributions into specific parameter packs. This allows them to achieve ``local dominance'' in targeted subspaces even while remaining a minority globally, effectively bypassing norm-based defenses that look for global outliers.

To resolve this geometric crisis, we propose \textbf{SafeSparse}, a consensus restoration framework that recouples sparsification and robustness through a dual-space calibration mechanism. Specifically, SafeSparse first enforces \textit{topological consensus} by utilizing Jaccard similarity to filter structural outliers induced by index poisoning. It then establishes \textit{semantic alignment} by mapping updates to a directional unit-sphere via sign-based vectors and employing density-based clustering. This approach effectively isolates the benign consensus cluster without relying on the flawed Euclidean magnitude, thereby neutralizing both structural and magnitude-invariant attacks.

Our contributions are summarized as follows: 
(1) We systematically identify and formalize the geometric dissonance problem in sparse FL, proving that standard robust aggregators possess a theoretical failure mode when inputs are sparse and orthogonal.
(2) We present \textit{SafeSparse}, the first defense framework explicitly tailored to reconcile the tension between communication efficiency and robust aggregation through mask-aware and sign-aware inspection. 
(3) We provide theoretical convergence guarantees for our method and demonstrate through extensive experiments that \textit{SafeSparse} significantly outperforms existing baselines, recovering global model accuracy in highly corrupted environments where traditional defenses fail.

\begin{figure}[htb]
\centering
\includegraphics[width=\linewidth]{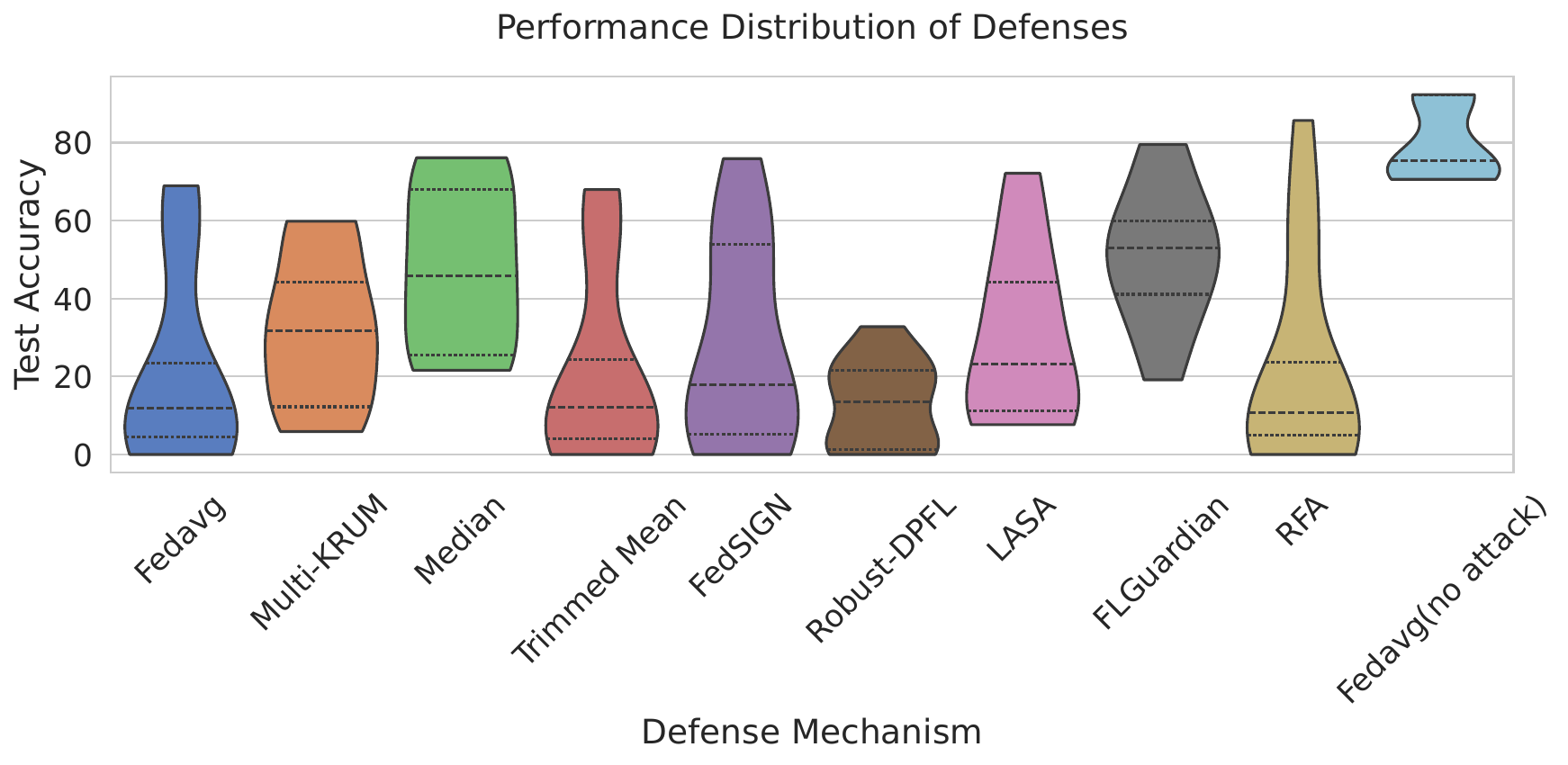}
% \vspace{-5mm}
\caption{\textbf{Empirical validation of poisoning vulnerability.} The violin plots illustrate the accuracy distribution across \textbf{12 diverse scenarios} (3 datasets $\times$ 4 attacks) as detailed in Section~\ref{baselines}. 
} 
\label{violin} 
% \vspace{-5mm}
\end{figure}

\begin{figure*}[htb]
\centering
\includegraphics[width=0.9\textwidth]{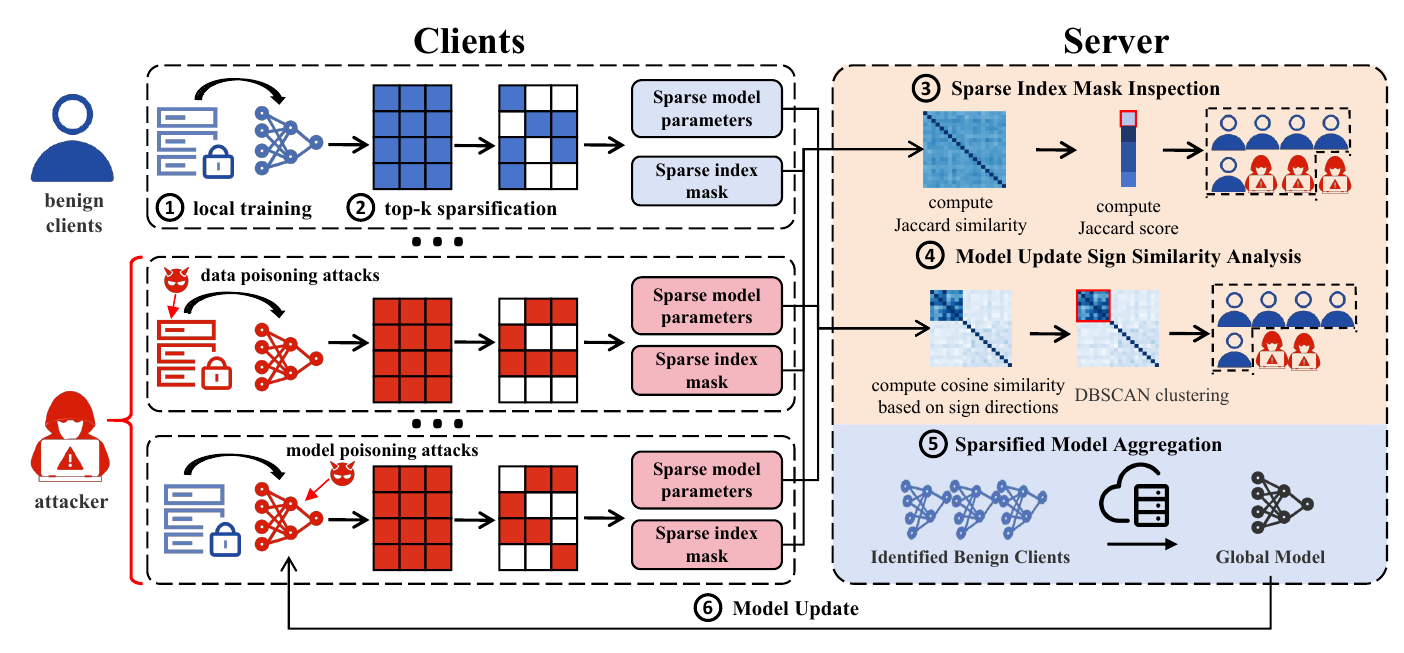}
% \vspace{-5mm}
\caption{
An illustration of the \textsc{SafeSparse} training process, incorporating client-side top-$k$ sparsification and server-side robust and sparsified aggregation.
It also highlights the two key components of SafeSparse: Sparse Index Mask Inspection and Model Update Sign Similarity Analysis. 
The red sections indicate attackers, who may carry out data poisoning attacks (such as label flipping attack) or model poisoning attacks (including IPM, Gaussian, and scaling attacks) to degrade the performance of the global model.
} 
\label{overview} 
\vspace{-5mm}
\end{figure*}

\section{Vulnerabilities in Sparsified FL}
\label{sec:sm}

Traditional robust FL often assumes that malicious updates are ``outliers'' in the full parameter space. However, we identify a fundamental shift in the attack surface when sparsification-based communication-efficiency method is employed. 

% In sparsified FL, participants only upload a fraction of coordinates $\mathcal{S} \subset \{1, \dots, d\}$. 
% This allows adversaries to perform subspace poisoning: by coordinating their sparse masks, multiple attackers can concentrate their malicious energy on a specific subset of parameters. Consequently, even if attackers are a minority globally, they can become a local majority within a specific sparse subspace.
Specifically, in the context of sparsified FL, we identify two distinct attack vectors.
(1) \textit{Index Poisoning}: Adversaries manipulate the sparse index masks to maximize the \textit{Malicious Contributor Ratio} ($f_p$) for specific target parameters, effectively hijacking the aggregation weight of critical neurons.
(2) \textit{Masked Value Manipulation}: Adversaries inject malicious perturbations (e.g., sign flipping) into the selected parameters while mimicking the sparsity patterns of benign clients to evade detection.
Unlike standard FL, where attackers are diluted globally, sparsification allows attackers to achieve local dominance in specific parameter packs.

\paragraph{Theoretical Analysis.}
Unlike dense FL, sparsified aggregation allows adversaries to concentrate their influence. We define attack effectiveness $\rho = \|W_G - W'_G\|_2^2$ as the squared Euclidean distance between the poisoned global model $W_G$ and an ideal benign model $W'_G$.

\begin{theorem}[Vulnerability under Sparsified Aggregation]
\label{thm:attack_rho}
Assume the malicious perturbation per parameter pack is bounded by $\mathbb{E}[\|W_a(p) - W'_G(p)\|_2] \le \epsilon$. The attack effectiveness $\rho$ is upper-bounded by:
\begin{equation}
    \rho \le \epsilon^2 \sum_{p=1}^P f_p^2,
\end{equation}
where $f_p = N_p^A / N_p$ is the malicious contributor ratio for parameter pack $p$.
\end{theorem}

Theorem~\ref{thm:attack_rho} reveals that the attack impact is determined by the pack-level ratio $f_p$ rather than the global attacker ratio in sparsified FL. By coordinating sparse masks, adversaries can drive $f_p \to 1$ for specific packs, exerting immense influence. \textit{SafeSparse} mitigates this by using Jaccard and Sign-based filtering to minimize $f_p$ across all packs. Detailed proof is provided in Appendix~C.

In addition to theoretical analysis, we experimentally verified the vulnerability.
As illustrated in Figure~\ref{violin}, we evaluate the global model accuracy across 12 diverse scenarios (comprising three datasets and four coordinated attacks). The probability density (represented by the width of the violins) and the quartiles (indicated by the dashed lines within) reveal a systemic performance collapse for traditional defenses. Specifically, mechanisms like \textit{Multi-KRUM} and \textit{RFA} exhibit extreme variance, with their interquartile ranges stretching significantly toward low-accuracy zones (often $<40\%$). 
This high instability empirically confirms Theorem~\ref{thm:attack_rho}: 
since these defenses fail to account for the mask-level distribution $f_p$, adversaries can effectively bypass geometric filters by concentrating malicious energy to become a ``local majority'' within coordinated sparse subspaces.

\section{SafeSparse Framework}
\label{sec:SafeSparse}

\subsection{Overview of SafeSparse}

% Existing sparsified communication efficiency approaches have been observed to obscure malicious modifications in poisoning attack scenarios, enabling adversarial clients to evade detection (demonstrated and discussed in Section~\ref{sec:vul}). 
% Consequently, existing defense mechanisms, originally designed for standard FL, become ineffective when applied to communication-efficient FL frameworks. These findings underscore the urgent need for security-aware sparsification techniques to ensure both robustness and communication efficiency in FL systems. 
% To address this challenge, the goal of the proposed SafeSparse framework is to enhance robustness against untargeted poisoning attacks in the context of sparsified, communication-efficient FL.

The goal of the proposed \textit{SafeSparse} framework is to enhance robustness against untargeted poisoning attacks in the context of sparsified, communication-efficient FL.
As illustrated in \figurename~\ref{overview}, \textit{SafeSparse} is built on a common communication-efficient FL paradigm, where each client communicates with the server by transmitting only the most important model parameters using sparsification techniques. 
Specifically, each client performs local training on its data, applies top-\(k\) sparsification, and uploads the sparse model parameters along with a sparse index mask to the server. 
The server then aggregates the received sparse model updates to form the global model based on the sparse index mask. 
% This sparsified model aggregation approach significantly reduces communication overhead, as only a fraction of the model parameters are exchanged.

To achieve the robustness guarantee and sparsification communication-efficiency in \textit{SafeSparse},
two key components are designed at the server side: sparse index mask inspection and model update sign similarity analysis. 
Specifically, the sparse index mask inspection module filters out clients whose model updates are inconsistent with the majority by measuring the overlap in parameter selections across clients using the Jaccard similarity metric. 
The model update sign similarity analysis module, another critical component of \textit{SafeSparse}, detects adversarial behavior patterns by grouping clients with similar update directions (i.e., the signs of parameter differences) using cosine similarity and applying density-based clustering to identify malicious clients. Clients identified as potentially malicious are excluded from the subsequent sparsified communication-efficient aggregation process, ensuring that only benign clients contribute to the final global model.

% In the following sections, we first introduce the robustness enhancement mechanism designed in SafeSparse, which addresses the challenges posed by poisoning attacks in sparsification-based communication within FL scenarios. 
% This mechanism defends against potential poisoning attacks while maintaining communication efficiency. 
% Subsequently, we integrate the proposed defense strategy into the standard sparsified, communication-efficient FL paradigm.

\paragraph{Security Model.}
In the \textit{SafeSparse} framework, we consider a Byzantine threat model in which the proportion of adversarial participating clients is assumed to remain below 50\% of the total clients. 
Each adversarial client is capable of launching poisoning attacks, aiming to manipulate the global model aggregation process by transmitting falsified model parameters during the FL process. 
This can be achieved either by directly manipulating the local model or by indirectly falsifying the local training samples. 
Consistent with the threat assumptions of existing poisoning attack studies, we assume that attackers do not have direct access to the local model updates or raw training data of benign participants. 
Furthermore, in this work, the primary adversarial objective is to conduct untargeted attacks designed to degrade the overall performance of the global model. 
It is important to note that privacy leakage threats, such as those caused by privacy inference attacks or gradient leakage attacks, are beyond the scope of this paper.

\subsection{Sparsified and Robust Aggregation}

% To mitigate poisoning attacks in sparse communication scenarios, we propose two key components: \textit{Sparse Index Mask Inspection} and \textit{Model Update Sign Similarity Analysis}. 
% These components are collaboratively designed to identify and filter out potential poisoning clients during the FL training process.
% The first component verifies whether the index mask of uploaded parameters adheres to expected patterns, ensuring the integrity of sparsified updates. 
% The second component detects adversarial behaviors by analyzing the similarity of model update directions, identifying patterns indicative of potential attacks. 
% We detail these components below.

% \subsubsection{Sparse Index Mask Inspection}
\subsubsection{Structural Integrity Verification via Jaccard Filtering}
Typical sparsified frameworks rely on \textit{sparse index masks}, which introduces the risk of \textit{Index Poisoning}. As shown in our theoretical derivation (Appendix~C), the success of an attack in sparse FL depends on the malicious contributor ratio $f_p$ within each parameter pack. To minimize this ratio, we enforce structural consensus.

\begin{figure}[htbp]
\centering
\includegraphics[width=\linewidth, trim=0 15pt 0 30pt, clip]{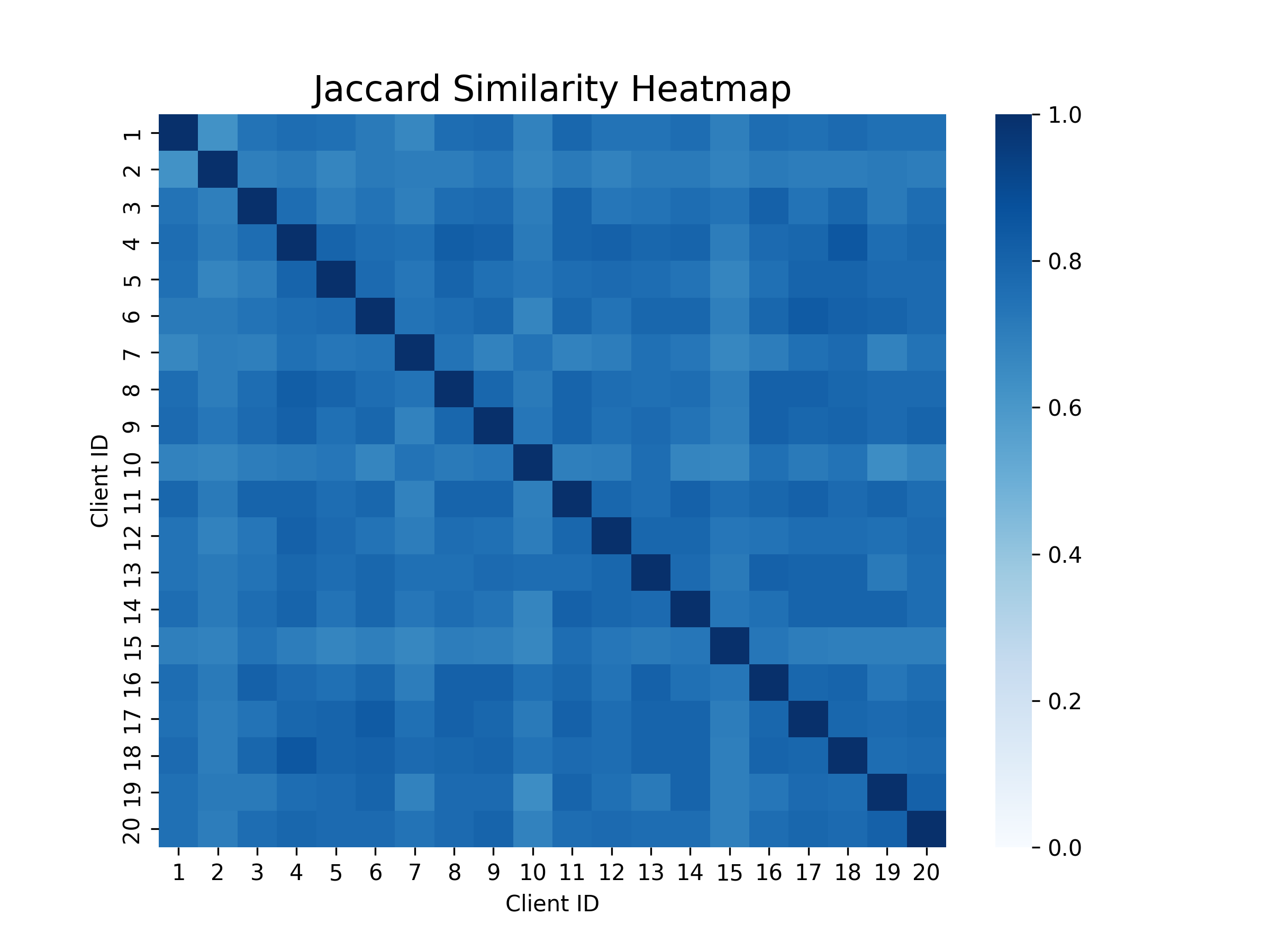}
\caption{The heatmap of the Jaccard similarity between the \textit{sparse index masks} of different clients on the CIFAR-10 dataset under the non-IID setting.} 
\label{jaccard_similarity} 
\vspace{-3mm}
\end{figure}

% Typical sparsified communication-efficient frameworks typically rely on \textit{sparse index masks}.
To further investigate their properties, we analyzed the structural characteristics of these masks and examined their consistency across clients. 
As shown in \figurename~\ref{jaccard_similarity}, our empirical analysis reveals that, even under non-IID training settings, a significant degree of overlap exists among the masks generated by different clients. 
Leveraging this observation, we propose a filtering mechanism to eliminate potential adversarial clients whose sparse index masks exhibit low overlap with others or form isolated groups of similarity. 
Typically, clients with highly distinct sparse index masks introduce a significant degree of uncertainty in detecting poisoning attacks, thereby hindering the establishment of a robust defense. 
By filtering out these outliers, our approach enhances the reliability of attack detection and ensures a more consistent aggregation process.

Specifically, we compute the Jaccard similarity between the sparse index masks of different clients. Formally, for each client \(i\), let \(M_i\) represent its sparse index mask, which contains the indices of the selected model parameters after top-\(k\) sparsification. The Jaccard similarity between two clients \(i\) and \(j\) is defined as:
\begin{equation}
J(M_i, M_j) = \frac{|M_i \cap M_j|}{|M_i \cup M_j|}
\end{equation}
where \(|M_i \cap M_j|\) represents the number of common selected indices between the two clients, and \(|M_i \cup M_j|\) represents the total number of unique selected indices. 

For each client, we compute its Jaccard score as the average Jaccard similarity with all other clients, defined as:
\begin{equation}
J_{\textsc{score}}(i) = \frac{1}{m-1} \sum_{j \neq i} J(M_i, M_j)
\label{eq:score}
\end{equation}
where \(m\) is the total number of participating clients. The Jaccard score quantifies the similarity of a client’s sparse index mask to those of the other clients in the system.

To identify anomalous clients, we define a threshold for filtering as follows:
\begin{equation}
J_{\textsc{threshold}} = \frac{J_{\textsc{max}} + J_{\textsc{min}}}{2} \cdot \beta
\label{eq:threshold}
\end{equation}
Here, \(J_{\text{max}}\) and \(J_{\text{min}}\) represent the highest and lowest Jaccard scores among all clients, respectively, and \(\beta\) is a manually tuned hyperparameter, with a default value of 0.6. Any client with a Jaccard score below this threshold is classified as an outlier and is excluded from the global model aggregation process.

% \subsubsection{Model Update Sign Similarity Analysis}
\subsubsection{Semantic Consistency via Sign-based Clustering}

\begin{figure}[htbp]
\centering
\includegraphics[width=\linewidth, trim=0 5pt 0 15pt, clip]{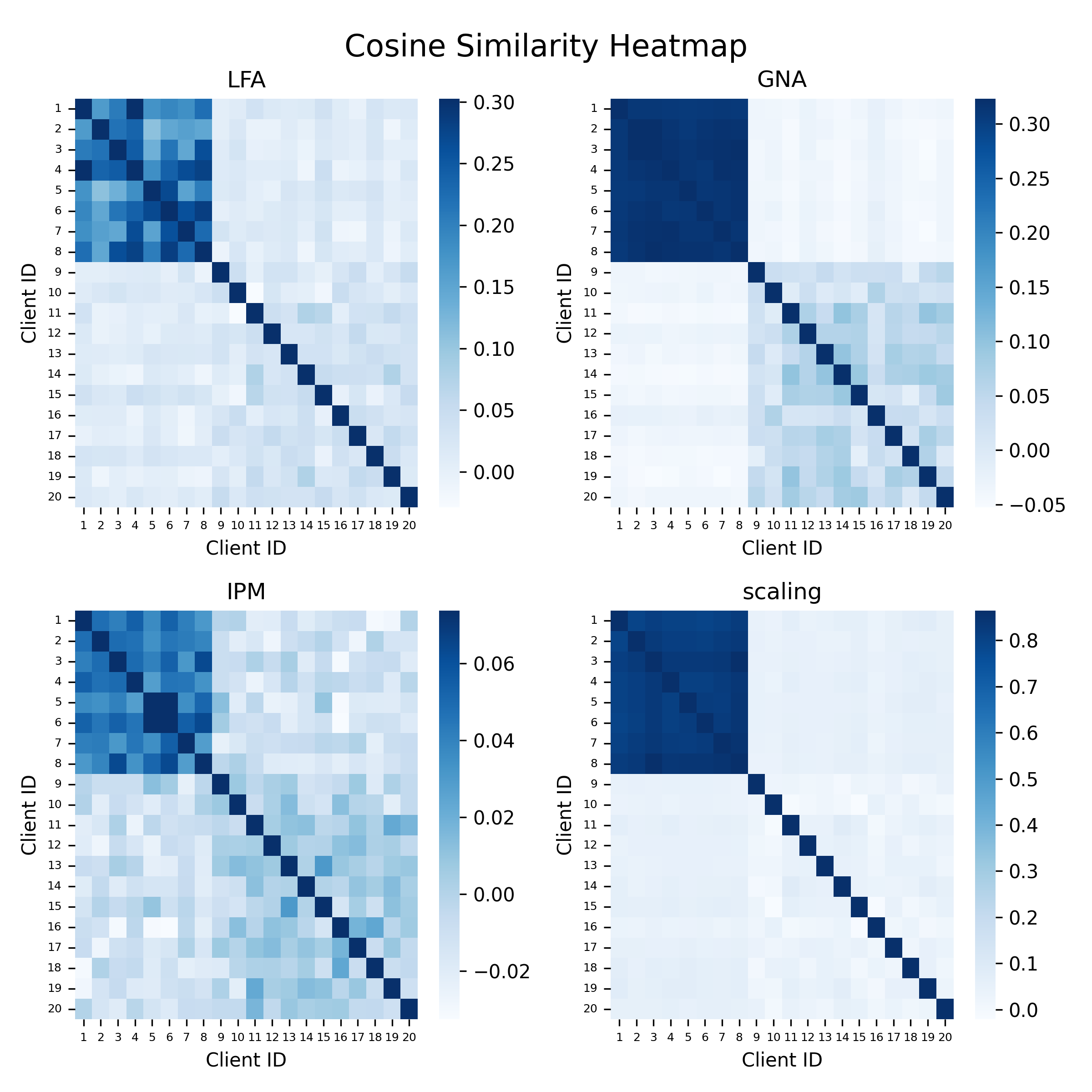}
% \vspace{-8mm}
\caption{The heatmap of sign cosine similarity among clients on the CIFAR-10 dataset under the non-IID setting. The first 8 clients are adversaries, while the remaining 12 are benign clients.} 
\label{cosine_similarity}
\vspace{-3mm}
\end{figure}

% Motivated by recent works \cite{guo2023fedsign}, we also leverage similarity measurements among clients as a fundamental strategy to mitigate poisoning attacks. To evaluate the effectiveness of this approach, we simulate four types of poisoning attacks in a sparsified communication scenario and analyze the similarity of model update signs among clients.

While Jaccard filtering ensures structural integrity, it cannot detect malicious values within a valid mask (e.g., Label Flipping or IPM). 
To address this, we examine the \textit{Semantic Consistency} of the updates. Since magnitude clipping is ineffective against scaling attacks in sparse regimes, we abstract the updates to their sign vectors $\text{Sign}(\Delta w_i)$. 
This transformation renders the defense robust to magnitude manipulation and focuses purely on the optimization direction.

As illustrated in \figurename~\ref{cosine_similarity}, the first eight clients represent malicious adversaries, while the remaining twelve are benign clients. The results demonstrate that the sign similarity among malicious clients is significantly higher than that among benign clients in the sparsified communication scenario. Furthermore, the similarity between malicious and benign clients remains low, further emphasizing the distinct update patterns introduced by adversarial attacks.
It is important to note that, due to the nature of sparsification, the sign similarity between any two clients can only be computed over the overlapping regions of their sparse index masks. 
In short, our observations reveal that the success of poisoning attacks typically relies on controlling multiple adversarial clients and injecting relatively consistent perturbations into the model updates.

Building upon these insights, we design a \textit{Model Update Sign Similarity Analysis} module to cluster potential malicious clients for detection. 
The process begins by computing the model parameter differences between the sparse model parameters uploaded by each client and the global model from the previous round, as $\Delta w_i = w_i - w_G$. 
These differences are then converted into their corresponding sign directions, defined as:
\begin{equation}
\begin{aligned}
\text{Sign}(\Delta w_i) &= \begin{cases} 
1, & \text{if } \Delta w_i > 0, \\
-1, & \text{if } \Delta w_i < 0,
\end{cases}
\end{aligned}
\label{eq:sign}
\end{equation}
where \(w_i\) represents the model parameters from client \(i\), and \(w_G\) denotes the global model parameters from the previous aggregation round.
It is important to note that parameters for which no client participated in the aggregation during the previous round are excluded from the similarity calculation. This exclusion arises because the server lacks the corresponding model parameters for these positions, as they are determined solely by the local models of the clients, which are independently set by each client.

Next, we compute the pairwise cosine similarity of model update signs between clients, considering only the overlapping portions of their sparse index masks. 
This restriction ensures that the comparison is performed on the same model elements between clients, thereby making the similarity measurement meaningful. 
The cosine similarity, \( \cos(\theta_{i,j}) \), is defined as follows:
\begin{equation}
\cos(\theta_{i,j}) = \frac{\langle \text{Sign}(\Delta w_i), \text{Sign}(\Delta w_j) \rangle}{\|\text{Sign}(\Delta w_i)\| \cdot \|\text{Sign}(\Delta w_j)\|},
\label{eq:cos}
\end{equation}
where \( \langle \cdot, \cdot \rangle \) denotes the dot product, and \( \|\cdot\| \) represents the Euclidean norm.

Subsequently, the cosine similarity is transformed into a distance metric using \( 1 - \cos(\theta_{i,j}) \), which quantifies the distance between the updates of two clients. The resulting distance matrix is then utilized as input to a clustering algorithm. Specifically, we employ the DBSCAN (Density-Based Spatial Clustering of Applications with Noise) algorithm \cite{ester1996density} within our framework, as detailed in Algorithm 1 in the Appendix B.1.

To determine the optimal value of \(\epsilon\), which defines the scan radius for core points in DBSCAN, we utilize the K-nearest neighbors (KNN) method. 
In this approach, \(\epsilon\) is set as the average distance to the \(n\)-th nearest neighbor, where \(n = N \cdot \gamma\). 
Here, \(N\) represents the total number of clients, and \(\gamma\) is a hyperparameter that controls the sensitivity of the clustering process. 
In practice, we set \(\gamma = 0.2\) to balance sensitivity to adversarial clients and false positives. Additionally, the minimum number of points required to form a core point in DBSCAN is set to \(N \cdot \gamma\), ensuring that each cluster contains a sufficient number of clients.
This clustering process enables the effective grouping of clients based on similar attack patterns, while clients exhibiting significantly different update behaviors are identified as outliers. 
The resulting clusters facilitate the detection and exclusion of malicious clients from the global aggregation process. Consequently, only benign clients are included in the final global model aggregation, thereby enhancing the robustness of the framework.

\paragraph{Sparsified Robust Aggregation.}
\label{sec:spar_agg}

% Model sparsification has emerged as a promising technique for improving communication efficiency in FL by reducing the number of transmitted model parameters during training \cite{sattler2019sparse,sattler2019robust}. 
% In traditional top-\(k\) sparsification setting, as proposed by Aji and Heafield \cite{aji2017sparse}, only the most significant model parameters are selected and transmitted at the scalar level, while the remaining parameters are set to zero. 
Building on the recent work of Yan et al. \cite{yan2024efficient}, \textit{SafeSparse} adopts a pack-level top-\(k\) sparsification method.
Unlike scalar-level sparsification, this method groups model parameters into small structured packs before applying the sparsification process. 
By operating at the pack level, this approach preserves the structured information within each group, providing a more efficient and cohesive representation of the model updates.

% Algorithm 2 (see Appendix B.2) outlines the sparsified aggregation framework designed for communication-efficient FL. 
Specifically, when the server receives sparsified model parameters $\{W_1,...,W_m\}$ and their sparse index masks $\{M_1,...,M_m\}$ from $m$ clients, it first resolves sparse index misalignment by identifying contributing clients $S_i=\{j|M_j(P_i)=1 \}$ for each packet $P_i$. 
This alignment ensures structural consistency across clients’ heterogeneous sparse representations. 
The aggregation weight for client $j$ is dynamically assigned as $\frac{|D_j|}{\sum_{k=1}^{m}|D_k|}$, proportional to its local dataset size $|D_j|$. For each $P_i$, the server calculates the total aggregation weight $A_{P_i} = \sum_{j \in S_i} \frac{|D_j|}{\sum_{k=1}^{m}|D_k|}$ to quantify cumulative contributions. 
Each model parameter $W_j(P_i)$ undergoes two sequential transformations: first scaled by its client-specific weight $W_j(P_i) \gets W_j(P_i) * \frac{|D_j|}{\sum_{k=1}^{m}|D_k|}$ then normalized by $A_{P_i}$ via $W_j(P_i) \gets W_j(P_i) / A_{P_i}$ to eliminate dimensional variance caused by differing contributor counts. The aggregated parameter $w_i^{agg}=\sum_{j\in S_i}W_j(P_i)$ is computed through weighted summation.
The detailed algorithm is provided in Appendix~B.2.

It is important to note that our aggregation employs a dynamic normalization factor $A_{P_i}$. Unlike standard FL averaging which divides by $N$, \textit{SafeSparse} normalizes by the count of \textit{valid contributing clients} for each pack. This prevents the ``gradient vanishing'' problem for rarely selected parameters and ensures that the global model maintains the correct scale of gradient descent, as guaranteed by our convergence proof in Appendix D.

\begin{figure*}[htb]
\centering 
\includegraphics[width=0.9\textwidth,trim=0 5pt 0 5pt,clip]{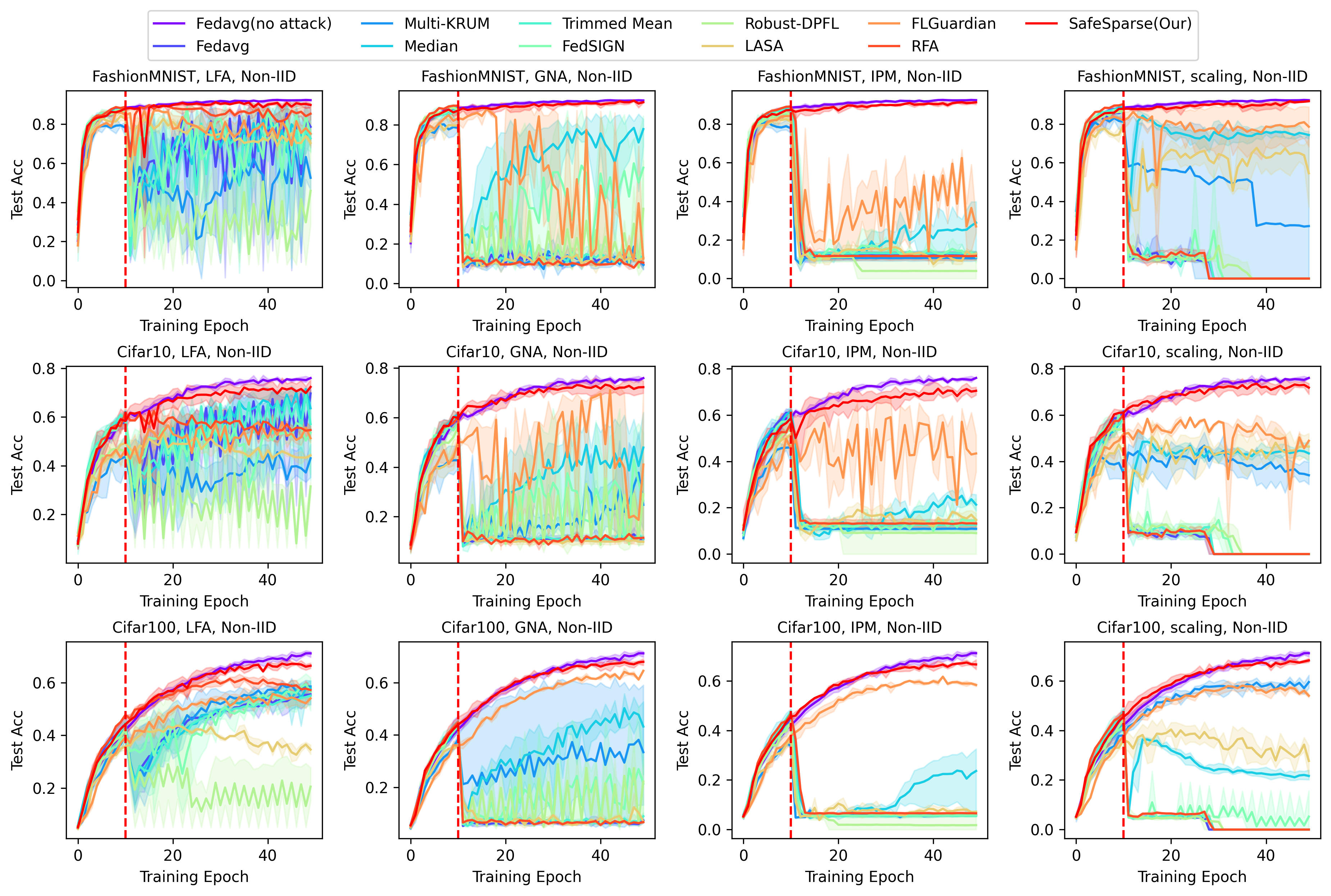}
\caption{Comparison of defense effectiveness across various approaches, evaluated on FashionMNIST, CIFAR-10, and CIFAR-100 under Label Flip Attack (LFA), Gaussian Noise Attack (GNA), Inner Product Manipulation (IPM), and Scaling Attack in a \textbf{NON-IID SETTING}. The attacker ratio is fixed at 0.4, and the top-$k$ sparsification rate is set to 0.5. Top-1 accuracy is used as the evaluation metric for FashionMNIST and CIFAR-10, while top-5 accuracy is adopted for CIFAR-100.}
\label{noniid}
\vspace{-3mm}
\end{figure*}

\section{Evaluations}
\label{sec:experiments}

\subsection{Convergence Analysis}
% In this section, we provide a  theoretical analysis to evaluate the robustness and stability of \textit{SafeSparse}. 
% We provide a formal convergence guarantee under adversarial conditions.
We further establish the convergence of \textit{SafeSparse} under standard assumptions and adversarial conditions (L-smoothness, bounded variance, and Top-$k$ properties, see Appendix~D.1).

\begin{theorem}[Convergence of SafeSparse]
\label{thm:convergence}
Let $\alpha = k/d$ be the fraction of retained parameters. Under a $\mu$-strongly convex objective and learning rate $\eta_t = \frac{8}{\mu(t+a)}$, the global model $W_G^T$ after $T$ rounds satisfies:
\begin{equation}
\mathbb{E}[\|W_G^T - W^*\|^2] \le \frac{D}{T+a} + \underbrace{\mathcal{O}\left(\frac{(1-\alpha)G^2}{\alpha^2}\right)}_{\text{Sparsification Error}} + \underbrace{\mathcal{O}\left(\frac{\rho}{\mu^2}\right)}_{\text{Residual Attack Impact}},
\end{equation}
where $D$ is a constant, $G^2$ represents data heterogeneity, and $\rho$ is the residual attack effectiveness defined in Theorem~\ref{thm:attack_rho}.
\end{theorem}

Theorem~\ref{thm:convergence} demonstrates that \textit{SafeSparse} converges to an error ball whose radius is explicitly controlled by the sparsity ratio $\alpha$ and the filtering effectiveness. 
By filtering outliers, \textit{SafeSparse} restricts $\rho$, ensuring the model converges even in highly adversarial sparse environments. 
Detailed proof is deferred to Appendix~D.

\subsection{Default Experimental Settings}

% Our proposed \textit{SafeSparse} is implemented using PFLlib \cite{zhang2023pfllib}, an open-source framework for federated learning (FL).
In the following experiments, unless stated otherwise, the default FL training setup includes 20 clients, all participating in each training round and performing one local training epoch. The attacker ratio is set to 0.4, with poisoning attacks starting at the 10th round and continuing until training ends. The batch size is 64, and the Adam optimizer is used with a learning rate of 0.005.

We evaluate \textit{SafeSparse} under two common data partitioning settings in FL: IID and Non-IID. In the IID setting, the dataset is evenly distributed among clients, with each client receiving an equal share of samples from all classes to ensure uniformity. Conversely, the Non-IID setting simulates statistical heterogeneity using a Dirichlet distribution-based method \cite{minka2000estimating}, where a concentration parameter $\alpha$ determines the level of data heterogeneity. For our experiments, we set $\alpha=1$ to achieve a moderate level of data heterogeneity among clients.

\subsection{Datasets and Baselines}
\label{baselines}

To evaluate the effectiveness and performance of \textit{SafeSparse}, we conduct several experiments, consistent with the methodologies employed in existing related works. Specifically, we train a ResNet-20 model \cite{he2016deep} using SafeSparse and compare it against related baselines on three public datasets: FashionMNIST \cite{xiao2017fashion}, CIFAR-10 \cite{krizhevsky2009learningCifar}, and CIFAR-100 \cite{krizhevsky2009learningCifar}.

\textit{SafeSparse} is compared against a comprehensive set of baselines, including: FedAVG \cite{mcmahan2017communication}, Multi-KRUM \cite{blanchard2017machine}, Median \cite{yin2018byzantine}, Trimmed Mean \cite{yin2018byzantine}, FedSign \cite{guo2023fedsign}, Robust-DPFL \cite{qi2024towards}, LASA \cite{xu2025achieving}, FLGuardian \cite{zhou2025flguardian}, and RFA \cite{pillutla2022robust}. The full technical descriptions and mathematical formulations for each baseline are detailed in Appendix~E.1.

To evaluate \textit{SafeSparse}'s resilience to poisoning attacks, we conducted experiments on both IID and non-IID datasets under four common poisoning attack scenarios: Label Flip Attack (LFA) \cite{fang2020local}, Gaussian Noise Attack (GNA) \cite{fang2020local}, Inner Product Manipulation (IPM) \cite{xie2020fall}, and Scaling Attack \cite{bagdasaryan2020backdoor}. Detailed definitions and the implementation mechanisms for these four attack strategies are further elaborated in Appendix~E.1.

% \begin{figure*}[h]
% \centering
% \includegraphics[width=0.9\textwidth,trim=0 5pt 0 5pt,clip]{images/beta_Compare_PerDataset.png}
% \caption{Comparison of defense effectiveness across different $\beta$ settings, evaluated on CIFAR-100 under LFA, GNA, IPM attack, and scaling attack.}
% \label{beta}
% \end{figure*}

% \begin{figure*}[h]
% \centering 
% \includegraphics[width=0.9\textwidth,trim=0 5pt 0 5pt,clip]{images/gama_Compare_PerDataset.png}
% \caption{Comparison of defense effectiveness across different $\gamma$ settings, evaluated on CIFAR-100 under LFA, GNA, IPM attack, and scaling attack.}
% \label{gamma}
% \end{figure*}

\begin{table*}[h]
\centering
\caption{Comparison of defense effectiveness across different $\beta$ settings (Non-IID Distribution).}
\label{table:beta_non_iid}
\small
\begin{tabular}{llccccc}
\toprule
\textbf{Dataset} & \textbf{Attack} & \textbf{$\beta=0.20$} & \textbf{$\beta=0.40$} & \textbf{$\beta=0.60$} & \textbf{$\beta=0.80$} & \textbf{$\beta=1.00$} \\
\midrule
\multirow{4}{*}{FashionMNIST} 
 & LFA     & 0.9127 & \textbf{0.9258} & 0.9156 & 0.9000 & \textit{0.8705} \\
 & GNA     & 0.8990 & \textit{0.8594} & \textbf{0.9251} & 0.8904 & 0.9183 \\
 & Scaling & 0.8848 & 0.9092 & 0.9090 & \textbf{0.9161} & \textit{0.8749} \\
 & IPM     & 0.8988 & \textit{0.8438} & \textbf{0.9165} & 0.8572 & 0.8761 \\
\midrule
\multirow{4}{*}{CIFAR-10} 
 & LFA     & \textit{0.6399} & 0.6706 & \textbf{0.7472} & 0.6836 & 0.6504 \\
 & GNA     & \textbf{0.7243} & 0.6839 & 0.6537 & 0.7065 & \textit{0.6409} \\
 & Scaling & 0.6088 & 0.6888 & \textbf{0.7104} & 0.7019 & \textit{0.5281} \\
 & IPM     & \textbf{0.7151} & 0.6764 & 0.6887 & 0.6648 & \textit{0.6519} \\
\midrule
\multirow{4}{*}{CIFAR-100} 
 & LFA     & 0.6622 & \textbf{0.6780} & \textit{0.6534} & 0.6626 & 0.6301 \\
 & GNA     & 0.6775 & \textit{0.6639} & \textbf{0.6875} & 0.6691 & 0.6561 \\
 & Scaling & 0.6611 & \textbf{0.6784} & 0.6453 & 0.6658 & \textit{0.6009} \\
 & IPM     & \textbf{0.6723} & 0.6669 & 0.6671 & 0.6709 & \textit{0.5675} \\
\bottomrule
\end{tabular}
\vspace{-3mm}
\end{table*}

\begin{table*}[t]
\centering
\caption{Comparison of defense effectiveness across different $\gamma$ settings (Non-IID Distribution).}
\label{table:gamma_non_iid}
\small
\begin{tabular}{llccccccc >{\columncolor{gray!15}}c >{\columncolor{gray!15}}c}
\toprule
\textbf{Dataset} & \textbf{Attack} & \textbf{0.10} & \textbf{0.15} & \textbf{0.20} & \textbf{0.25} & \textbf{0.30} & \textbf{0.35} & \textbf{0.40} & \textbf{0.45} & \textbf{0.50} \\
\midrule
\multirow{4}{*}{FashionMNIST} 
 & LFA      & 0.9163 & \textbf{0.9223} & 0.9102 & 0.8777 & 0.9044 & 0.8520 & 0.9216 & 0.5873 & 0.7200 \\
 & GNA      & 0.7871 & 0.9110 & 0.9124 & 0.8300 & 0.8899 & \textbf{0.9242} & 0.8908 & 0.1069 & 0.1315 \\
 & Scaling  & 0.8746 & \textbf{0.9245} & 0.9061 & 0.8571 & 0.9107 & 0.8524 & 0.8912 & 0.1164 & 0.1365 \\
 & IPM      & \textbf{0.8935} & 0.7162 & 0.8907 & 0.7216 & 0.5947 & 0.8159 & 0.7978 & 0.1226 & 0.1071 \\
\midrule
\multirow{4}{*}{CIFAR-10} 
 & LFA      & \textbf{0.7376} & 0.5951 & 0.7104 & 0.6769 & 0.7297 & 0.6594 & 0.7169 & 0.5772 & 0.4913 \\
 & GNA      & 0.6547 & 0.6479 & 0.7072 & 0.6185 & 0.6431 & 0.6995 & \textbf{0.7251} & 0.1311 & 0.1241 \\
 & Scaling  & 0.7044 & 0.7198 & 0.6897 & 0.7158 & \textbf{0.7256} & 0.7173 & 0.6854 & 0.0901 & 0.1034 \\
 & IPM      & 0.1474 & \textbf{0.7043} & 0.6836 & 0.6756 & 0.6032 & 0.5381 & 0.4689 & 0.1370 & 0.1337 \\
\midrule
\multirow{4}{*}{CIFAR-100} 
 & LFA      & 0.6525 & 0.6621 & 0.6572 & \textbf{0.6711} & 0.6669 & 0.6596 & 0.6561 & 0.5099 & 0.5271 \\
 & GNA      & 0.6662 & 0.6599 & 0.6704 & 0.6786 & 0.6740 & \textbf{0.6840} & 0.6799 & 0.0589 & 0.0698 \\
 & Scaling  & \textbf{0.6810} & 0.6734 & 0.6684 & 0.6733 & 0.6547 & 0.6669 & 0.6599 & 0.0482 & 0.0572 \\
 & IPM      & 0.5944 & 0.6525 & \textbf{0.6791} & 0.5347 & 0.4912 & 0.6531 & 0.6146 & 0.0617 & 0.0586 \\
\bottomrule
\end{tabular}
\vspace{-3mm}
\end{table*}

\subsection{Attack Effectiveness in Sparsified FL}
\label{sec:vul}

As illustrated in \figurename~\ref{noniid}, we demonstrate the failure of most existing defense strategies against poisoning attacks in the context of sparsified communication-efficient FL. 
Different types of poisoning attacks commence at the 10th training epoch (marked by the red dashed line), and their impact on common sparsified aggregation is evident. Specifically, FedAvg fails to maintain stability, particularly under Scaling and Inner Product Manipulation (IPM) attacks, where the test accuracy drops drastically after the attack begins. A complementary evaluation under the IID setting is provided in Appendix~E.2, which further corroborates these vulnerabilities.

Existing Federated Learning (FL) defenses, including Multi-KRUM, Median, and Trimmed Mean, exhibit limited and inconsistent resilience against poisoning attacks. Their effectiveness is highly sensitive to the dataset and attack type; for instance, while some succeed on CIFAR-10, they fail catastrophically under IPM or scaling attacks on CIFAR-100. Even advanced methods like FedSIGN and Robust-DPFL struggle with high-intensity attacks on complex datasets. These vulnerabilities are exacerbated in sparsified FL. Traditional robust aggregators rely on full, aligned updates, which are disrupted by the inconsistent parameter subsets inherent in sparsification. Similarly, anomaly detectors like FedSIGN lose detection granularity as sparse representations obscure critical patterns. This allows adversaries to exploit sparse indices to evade detection, highlighting an urgent need for sparsity-aware robust mechanisms.

\subsection{Model Performance and Defense Effectiveness}

To evaluate our proposed \textit{SafeSparse}, we compare the model performance and defense effectiveness of \textit{SafeSparse} with those of baseline methods. 
As shown in \figurename~\ref{noniid}, SafeSparse maintains high test accuracy and stability even after the attack is introduced, outperforming other defenses in multiple adversarial settings. Specifically, Scaling and Inner Product Manipulation (IPM) attacks cause many baseline methods to collapse entirely, while \textit{SafeSparse} remains robust, preserving high accuracy. These results indicate the effectiveness of \textit{SafeSparse} against various poisoning attacks in adversarial federated learning environments. Detailed comparative results on IID datasets and the corresponding convergence curves are available in Appendix~E.2.

\subsection{Ablation Experiments}
\label{sec:ablation}

To further evaluate the robustness and adaptability of \textit{SafeSparse}, we conduct ablation experiments on its two key hyperparameters: the filtering threshold $\beta$ and the clustering sensitivity $\gamma$, as illustrated in Table~\ref{table:beta_non_iid} and Table~\ref{table:gamma_non_iid} , while the corresponding results for the IID distribution are provided in Appendix~E.3.

\paragraph{Impact of Filtering Threshold $\beta$.}
Filtering threshold controls the cutoff value for sparsity mask similarity when filtering out potentially poisoned clients. 
We experiment with five values: $\beta \in \{0.2, 0.4, 0.6, 0.8, 1.0\}$. As shown in our results, SafeSparse maintains effective defense performance under all settings, indicating its robustness to $\beta$ variations. However, higher values of $\beta$ (e.g., 0.8 or 1.0) lead to over-filtering, where benign clients are mistakenly excluded from training, reducing model diversity and performance. On the other hand, lower values retain more benign clients while still effectively filtering out attackers. We adopt $\beta=0.6$ in our main experiments as it strikes a good balance between defense strength and benign client retention.

\paragraph{Impact of Clustering Sensitivity Parameter $\gamma$.}
Clustering sensitivity parameter influences the neighborhood size for DBSCAN-based client clustering. We vary $\gamma$ from 0.1 to 0.5 in increments of 0.05. The results reveal that excessively high values (e.g., $\gamma > 0.4$) cause the clustering step to fail in separating attackers from benign clients, leading to ineffective defense.

\paragraph{Impact of Attacker Ratio and Top-$k$ Ratio.}
\textit{SafeSparse} was evaluated with attacker ratios from 0.2 to 0.4 on CIFAR-100. Results show that test accuracy remains stable regardless of the increasing proportion of malicious clients, confirming the robustness of our defense. See Appendix~E.4 for detailed performance curves.
We also tested sparsification ratios between 0.4 and 0.6. \textit{SafeSparse} maintains strong resilience across all levels, though lower ratios may slightly affect attacker detection stability. Further experimental data and analysis are provided in Appendix~E.5.

\section{Conclusion}
\label{sec:conclusion}

In this work, we investigated the security vulnerabilities of federated learning in sparsification based communication-efficient scenarios under poisoning attacks and demonstrated the limitations of existing defense mechanisms in addressing these threats. To overcome these challenges, we proposed SafeSparse, a robust defense framework to effectively detect and mitigate adversarial clients. Through extensive experiments, we demonstrated that SafeSparse significantly enhances the robustness of FL in sparsification scenarios while preserving communication efficiency. These findings underscore the importance of developing security-aware sparsification strategies, providing a foundation for future research on strengthening the security of communication-efficient FL systems.

% \bibliographystyle{abbrv}
% \bibliography{refs}

\bibliographystyle{named}
\bibliography{refs}

@article{xiao2017fashion,
  title={Fashion-mnist: a novel image dataset for benchmarking machine learning algorithms},
  author={Xiao, Han and Rasul, Kashif and Vollgraf, Roland},
  journal={arXiv preprint arXiv:1708.07747},
  year={2017}
}

@article{krizhevsky2009learningCifar,
  title={Learning multiple layers of features from tiny images},
  author={Krizhevsky, Alex and Hinton, Geoffrey and others},
  year={2009},
  publisher={Toronto, ON, Canada}
}

@misc{minka2000estimating,
  title={Estimating a Dirichlet distribution},
  author={Minka, Thomas},
  year={2000},
  publisher={Technical report, MIT}
}

@inproceedings{he2016deep,
  title={Deep residual learning for image recognition},
  author={He, Kaiming and Zhang, Xiangyu and Ren, Shaoqing and Sun, Jian},
  booktitle={Proceedings of the IEEE conference on computer vision and pattern recognition},
  pages={770--778},
  year={2016}
}

@inproceedings{fang2020local,
  title={Local model poisoning attacks to $\{$Byzantine-Robust$\}$ federated learning},
  author={Fang, Minghong and Cao, Xiaoyu and Jia, Jinyuan and Gong, Neil},
  booktitle={29th USENIX security symposium (USENIX Security 20)},
  pages={1605--1622},
  year={2020}
}

@inproceedings{xie2020fall,
  title={Fall of empires: Breaking byzantine-tolerant sgd by inner product manipulation},
  author={Xie, Cong and Koyejo, Oluwasanmi and Gupta, Indranil},
  booktitle={Uncertainty in Artificial Intelligence},
  pages={261--270},
  year={2020},
  organization={PMLR}
}

@inproceedings{bagdasaryan2020backdoor,
  title={How to backdoor federated learning},
  author={Bagdasaryan, Eugene and Veit, Andreas and Hua, Yiqing and Estrin, Deborah and Shmatikov, Vitaly},
  booktitle={International conference on artificial intelligence and statistics},
  pages={2938--2948},
  year={2020},
  organization={PMLR}
}

@inproceedings{mcmahan2017communication,
  title={Communication-efficient learning of deep networks from decentralized data},
  author={McMahan, Brendan and Moore, Eider and Ramage, Daniel and Hampson, Seth and y Arcas, Blaise Aguera},
  booktitle={Artificial intelligence and statistics},
  pages={1273--1282},
  year={2017},
  organization={PMLR}
}

@article{blanchard2017machine,
  title={Machine learning with adversaries: Byzantine tolerant gradient descent},
  author={Blanchard, Peva and El Mhamdi, El Mahdi and Guerraoui, Rachid and Stainer, Julien},
  journal={Advances in neural information processing systems},
  volume={30},
  year={2017}
}

@inproceedings{yin2018byzantine,
  title={Byzantine-robust distributed learning: Towards optimal statistical rates},
  author={Yin, Dong and Chen, Yudong and Kannan, Ramchandran and Bartlett, Peter},
  booktitle={International conference on machine learning},
  pages={5650--5659},
  year={2018},
  organization={Pmlr}
}

@article{guo2023fedsign,
  title={FedSIGN: A sign-based federated learning framework with privacy and robustness guarantees},
  author={Guo, Zhenyuan and Xu, Lei and Zhu, Liehuang},
  journal={Computers \& Security},
  volume={135},
  pages={103474},
  year={2023},
  publisher={Elsevier}
}

@inproceedings{qi2024towards,
  title={Towards the Robustness of Differentially Private Federated Learning},
  author={Qi, Tao and Wang, Huili and Huang, Yongfeng},
  booktitle={Proceedings of the AAAI Conference on Artificial Intelligence},
  volume={38},
  number={18},
  pages={19911--19919},
  year={2024}
}

@inproceedings{guan2023enabling,
  title={Enabling Communication-Efficient Federated Learning via Distributed Compressed Sensing},
  author={Guan, Yixuan and Liu, Xuefeng and Ren, Tao and Niu, Jianwei},
  booktitle={IEEE Conference on Computer Communications},
  pages={1--10},
  year={2023},
  organization={IEEE}
}

@inproceedings{honig2022dadaquant,
  title={DAdaQuant: Doubly-adaptive quantization for communication-efficient federated learning},
  author={H{\"o}nig, Robert and Zhao, Yiren and Mullins, Robert},
  booktitle={International Conference on Machine Learning},
  pages={8852--8866},
  year={2022},
  organization={PMLR}
}

@inproceedings{yan2024efficient,
  title={Efficient and straggler-resistant homomorphic encryption for heterogeneous federated learning},
  author={Yan, Nan and Li, Yuqing and Chen, Jing and Wang, Xiong and Hong, Jianan and He, Kun and Wang, Wei},
  booktitle={IEEE Conference on Computer Communications},
  pages={791--800},
  year={2024},
  organization={IEEE}
}

@article{wang2023sparsfa,
  title={SparSFA: Towards robust and communication-efficient peer-to-peer federated learning},
  author={Wang, Han and Mu{\~n}oz-Gonz{\'a}lez, Luis and Hameed, Muhammad Zaid and Eklund, David and Raza, Shahid},
  journal={Computers \& Security},
  volume={129},
  pages={103182},
  year={2023},
  publisher={Elsevier}
}

@inproceedings{cao2024efficient,
  title={Efficient Multi-Task Asynchronous Federated Learning in Edge Computing},
  author={Cao, Xinyuan and Ouyang, Tao and Zhao, Kongyange and Li, Yousheng and Chen, Xu},
  booktitle={2024 IEEE/ACM 32nd International Symposium on Quality of Service (IWQoS)},
  pages={1--10},
  year={2024},
  organization={IEEE}
}

@inproceedings{zang2024efficient,
  title={Efficient Asynchronous Federated Learning with Prospective Momentum Aggregation and Fine-Grained Correction},
  author={Zang, Yu and Xue, Zhe and Ou, Shilong and Chu, Lingyang and Du, Junping and Long, Yunfei},
  booktitle={Proceedings of the AAAI Conference on Artificial Intelligence},
  volume={38},
  number={15},
  pages={16642--16650},
  year={2024}
}

@article{li2023fedcompass,
  title={FedCompass: efficient cross-silo federated learning on heterogeneous client devices using a computing power aware scheduler},
  author={Li, Zilinghan and Chaturvedi, Pranshu and He, Shilan and Chen, Han and Singh, Gagandeep and Kindratenko, Volodymyr and Huerta, Eliu A and Kim, Kibaek and Madduri, Ravi},
  journal={arXiv preprint arXiv:2309.14675},
  year={2023}
}

@article{shi2019understanding,
  title={Understanding top-k sparsification in distributed deep learning},
  author={Shi, Shaohuai and Chu, Xiaowen and Cheung, Ka Chun and See, Simon},
  journal={arXiv preprint arXiv:1911.08772},
  year={2019}
}

@article{aji2017sparse,
  title={Sparse communication for distributed gradient descent},
  author={Aji, Alham Fikri and Heafield, Kenneth},
  journal={arXiv preprint arXiv:1704.05021},
  year={2017}
}

@article{sattler2019robust,
  title={Robust and communication-efficient federated learning from non-iid data},
  author={Sattler, Felix and Wiedemann, Simon and M{\"u}ller, Klaus-Robert and Samek, Wojciech},
  journal={IEEE transactions on neural networks and learning systems},
  volume={31},
  number={9},
  pages={3400--3413},
  year={2019},
  publisher={IEEE}
}

@inproceedings{sattler2019sparse,
  title={Sparse binary compression: Towards distributed deep learning with minimal communication},
  author={Sattler, Felix and Wiedemann, Simon and M{\"u}ller, Klaus-Robert and Samek, Wojciech},
  booktitle={2019 International Joint Conference on Neural Networks (IJCNN)},
  pages={1--8},
  year={2019},
  organization={IEEE}
}

@article{kairouz2021advances,
  title={Advances and open problems in federated learning},
  author={Kairouz, Peter and McMahan, H Brendan and Avent, Brendan and Bellet, Aur{\'e}lien and Bennis, Mehdi and Bhagoji, Arjun Nitin and Bonawitz, Kallista and Charles, Zachary and Cormode, Graham and Cummings, Rachel and others},
  journal={Foundations and trends{\textregistered} in machine learning},
  volume={14},
  number={1--2},
  pages={1--210},
  year={2021},
  publisher={Now Publishers, Inc.}
}

@inproceedings{han2020adaptive,
  title={Adaptive gradient sparsification for efficient federated learning: An online learning approach},
  author={Han, Pengchao and Wang, Shiqiang and Leung, Kin K},
  booktitle={2020 IEEE 40th international conference on distributed computing systems (ICDCS)},
  pages={300--310},
  year={2020},
  organization={IEEE}
}

@inproceedings{ester1996density,
  title={A density-based algorithm for discovering clusters in large spatial databases with noise},
  author={Ester, Martin and Kriegel, Hans-Peter and Sander, J{\"o}rg and Xu, Xiaowei and others},
  booktitle={kdd},
  volume={96},
  number={34},
  pages={226--231},
  year={1996}
}

@inproceedings{xu2025achieving,
  title={Achieving byzantine-resilient federated learning via layer-adaptive sparsified model aggregation},
  author={Xu, Jiahao and Zhang, Zikai and Hu, Rui},
  booktitle={2025 IEEE/CVF Winter Conference on Applications of Computer Vision (WACV)},
  pages={1508--1517},
  year={2025},
  organization={IEEE}
}

@article{zhou2025flguardian,
  title={FLGuardian: Defending against Model Poisoning Attacks via Fine-grained Detection in Federated Learning},
  author={Zhou, Xingjie and Chen, Xianzhang and Liu, Shukan and Fan, Xuehong and Sun, Qiao and Chen, Lin and Qiu, Meikang and Xiang, Tao},
  journal={IEEE Transactions on Information Forensics and Security},
  year={2025},
  publisher={IEEE}
}

@article{pillutla2022robust,
  title={Robust aggregation for federated learning},
  author={Pillutla, Krishna and Kakade, Sham M and Harchaoui, Zaid},
  journal={IEEE Transactions on Signal Processing},
  volume={70},
  pages={1142--1154},
  year={2022},
  publisher={IEEE}
}

@article{jiang2025towards,
  title={Towards efficient and certified recovery from poisoning attacks in federated learning},
  author={Jiang, Yu and Shen, Jiyuan and Liu, Ziyao and Tan, Chee Wei and Lam, Kwok-Yan},
  journal={IEEE Transactions on Information Forensics and Security},
  year={2025},
  publisher={IEEE}
}

@article{yang2024roseagg,
  title={Roseagg: Robust defense against targeted collusion attacks in federated learning},
  author={Yang, He and Xi, Wei and Shen, Yuhao and Wu, Canhui and Zhao, Jizhong},
  journal={IEEE Transactions on Information Forensics and Security},
  volume={19},
  pages={2951--2966},
  year={2024},
  publisher={IEEE}
}

@article{mu2024fedpta,
  title={FedPTA: Prior-based tensor approximation for detecting malicious clients in federated learning},
  author={Mu, Xutong and Cheng, Ke and Liu, Teng and Zhang, Tao and Geng, Xueli and Shen, Yulong},
  journal={IEEE Transactions on Information Forensics and Security},
  year={2024},
  publisher={IEEE}
}

@article{zhang2024fltracer,
  title={Fltracer: Accurate poisoning attack provenance in federated learning},
  author={Zhang, Xinyu and Liu, Qingyu and Ba, Zhongjie and Hong, Yuan and Zheng, Tianhang and Lin, Feng and Lu, Li and Ren, Kui},
  journal={IEEE Transactions on Information Forensics and Security},
  volume={19},
  pages={9534--9549},
  year={2024},
  publisher={IEEE}
}

\appendix

\section{Related Work}
\label{sec:related}

\subsection{Communication-Efficient FL}

Enhancing communication efficiency is critical for improving overall performance of FL systems. 
The substantial volume of data exchanged between clients and servers imposes significant challenges,
including increased latency, computational overhead, and excessive bandwidth consumption, which can hinder the system's real-time capabilities.
To address these issues, techniques such as sparsification\cite{yan2024efficient,wang2023sparsfa}, quantization\cite{guan2023enabling,honig2022dadaquant}, and asynchronous communication\cite{cao2024efficient,zang2024efficient,li2023fedcompass} have been proposed. 
These methods effectively reduce communication overhead, enabling FL systems to operate efficiently in resource-constrained environments while preserving model performance.

Sparsification methods have shown greater promise compared to alternative techniques, as quantization is not suitable for large-scale models or low-bandwidth network environments \cite{shi2019understanding}, and the staleness issue introduced by asynchronous aggregation can affect model convergence\cite{li2023fedcompass}.
Sparsification aims to reduce communication overhead by selecting only a subset of important gradients or model parameters for transmission during each training round, while less significant information is either ignored or compressed. 
Aji et al.\cite{aji2017sparse} proposed a method that selects the top-k components with the highest absolute values from the gradients and designates the remaining components as residuals. 
Subsequent studies \cite{sattler2019robust} have  demonstrated that Top-\(k\) sparsification is less sensitive to the impact of non-IID data in FL.
In the context of peer-to-peer federated learning(P2PFL), Wang et al.\cite{wang2023sparsfa} introduced a momentum-based Top-\(k\) sparsification method, termed SparSFA, to further enhance communication efficiency.

\subsection{Model Poisoning Attacks and Defense Strategies in FL}

The distributed architecture of FL makes it particularly vulnerable to poisoning attacks, where adversarial clients manipulate local updates to compromise the performance of the global model. 
% This study focuses on untargeted poisoning attacks, which aim to degrade the model's overall performance across all classes without targeting specific outputs.
Several types of untargeted poisoning attacks have been proposed in the literature. Label flip attack\cite{fang2020local} manipulates local training data by flipping class labels, thereby injecting mislabeled samples that degrade the model’s generalization. 
Gaussian noise attack\cite{fang2020local} generates model updates by adding random noise, leading to instability and reduced convergence quality. 
Inner product manipulation\cite{xie2020fall} strategically crafts adversarial updates to maximize the inner product with benign gradients, thereby amplifying their negative impact on model convergence. 
Scaling attacks\cite{bagdasaryan2020backdoor} introduce malicious updates that are disproportionately scaled, either by inflating or deflating gradient magnitudes, which disrupts the federated aggregation process and significantly impairs the effectiveness of the global model. These attacks demonstrate the vulnerability of FL to malicious participants, highlighting the need for robust defense mechanisms to mitigate their effects.

To mitigate the impact of poisoning attacks in FL, various defense strategies have been proposed. These strategies can be broadly categorized into robust aggregation\cite{yin2018byzantine,pillutla2022robust} rules and anomaly detection techniques\cite{blanchard2017machine,zhou2025flguardian,xu2025achieving,jiang2025towards,yang2024roseagg,mu2024fedpta,zhang2024fltracer}. 
Robust aggregation methods, such as Median \cite{yin2018byzantine} and Trimmed Mean \cite{yin2018byzantine}, aim to improve the resilience of the global model by mitigating the impact of malicious updates during the aggregation process.
Median aggregation selects the component-wise median of all client updates, making it robust to outliers and adversarial manipulations. Trimmed Mean excludes a fraction of the largest and smallest update values before computing the mean, effectively mitigating the impact of extreme deviations introduced by adversaries. Anomaly detection techniques take a different approach by identifying and filtering out malicious clients before aggregation. 
In addition, Krum\cite{blanchard2017machine}, a widely used anomaly detection-based method, selects a single client update that is closest to its neighbors in Euclidean space, assuming that the majority of clients are honest. 
By prioritizing updates that exhibit consistency with the majority, Krum effectively reduces the impact of outliers. 
However, these defense methods were primarily designed for standard FL and have not yet addressed attacker detection in sparse training scenarios.

\section{SafeSparse Algorithm}

\subsection{Poisoning Client Filtering}
\label{sec:pcf_appendix}

\begin{algorithm}[htb]  
    \caption{Poisoning Client Filtering}  
    \label{alg:pcf} 
    
    \SetKwInput{KwInput}{Input}
    \SetKwInput{KwOutput}{Output}
    \SetKwProg{Fn}{function}{}{}

    \KwInput{
        Sparse model $W=\{W_1, \dots, W_m\}$, sparse index masks $M=\{M_1, \dots, M_m\}$ of $m$ clients $C$; filtering threshold $\beta$ and clustering sensitivity $\gamma$. 
    }
    \KwOutput{
        Identified benign clients $C_\textsc{benign}$.
    }
    \Fn{poison\_filtering($C, W, M, \beta, \gamma$)}{
        Initialize $C_\textsc{benign} \gets C$ \;
        \ForEach{client $i$ in $C$}{
            $J_\textsc{score}(i) \gets \frac{1}{m-1} \sum_{j \neq i} J(M_i, M_j)$\;
        }
        Set filtering threshold $J_\textsc{threshold} = \frac{J_\textsc{max} + J_\textsc{min}}{2} * \beta$\;
        \ForEach{client $i$ in $C$}{
            \If{$J_\textsc{score}(i) < J_\textsc{threshold}$}{
                Exclude client $i$ from $C_\textsc{benign}$
            }
        }
    
        \ForEach{remaining client $i$ in $C_\textsc{benign}$}{
            Convert $\Delta w_i$ to sign direction by Equation~\ref{eq:sign}
        }
        Compute cosine similarity $cos(\theta_{i,j})$ by Equation~\ref{eq:cos}\;
        Compute distance matrix $\mathcal{D} = 1 - cos(\theta_{i,j})$\;
        Compute neighbor count $n = m * \gamma$\;
        Set $\epsilon$ as average distance to $n$-th nearest neighbor\;
        Apply DBSCAN clustering on $\mathcal{D}$ with $n, \epsilon$\;
        Exclude grouped clients from $C_\textsc{benign}$\;
        
        \Return $C_\textsc{benign}$
    }

\end{algorithm}

Algorithm~\ref{alg:pcf} implements the poisoning mitigation mechanism of SafeSparse. The process begins with an analysis of the structural consistency of sparse index masks using the Jaccard similarity metric. 
For each given client \(i\), its Jaccard score, \(J_{\textsc{score}}(i)\), is computed. 
A threshold \(J_\textsc{threshold}\) is then applied to filter out clients with \(J_{\textsc{score}}(i) < J_\textsc{threshold}\), effectively removing outliers that exhibit abnormal sparsification patterns. 
For the remaining clients, parameter updates are transformed into binary sign directions, emphasizing update polarity rather than magnitude. 
Subsequently, the pairwise directional similarity \( \cos(\theta_{i,j}) \) is calculated. 
The resulting distance matrix, \(1 - \cos(\theta_{i,j})\), is utilized as input to the DBSCAN clustering algorithm with adaptive parameters. 
Specifically, \(\epsilon\) is set as the average distance to the \(n\)-th nearest neighbor, where \(n = m \cdot \gamma\). Clients grouped into dense clusters by DBSCAN are identified as coordinated attackers and excluded from the set of benign clients.

\subsection{Sparsified Robust Model Aggregation}

In the sparsified, communication-efficient aggregation process of SafeSparse, each client first flattens its local model parameters and organizes them into structured groups for efficient packing and transmission. Based on the specified sparsification ratio and the characteristics of its local model, the client computes a sparse index mask to identify which parameter packs will be retained. Specifically, the client determines a sparsification threshold and assigns a mask value of 1 to packs whose aggregated values exceed this threshold. The final transmission to the server includes two components: the sparsified model parameters and a sparse index mask, represented with 1-bit per entry, indicating the retained packs. This structured sparsification approach effectively preserves critical information while significantly reducing communication overhead. Upon receiving these sparse updates, the server executes the robust aggregation procedure to filter potential malicious contributions and reconstruct the global model, as detailed in Algorithm~\ref{alg:sa}.

\begin{algorithm}[htb]  
    \caption{Sparsified Robust Model Aggregation}  
    \label{alg:sa} 
    \SetKwInput{KwInput}{Input}
    \SetKwInput{KwOutput}{Output}
    \SetKwProg{Fn}{function}{}{}

    \KwInput{Sparse model $W=\{W_1, \dots, W_m\}$, sparse index masks 
      $M=\{M_1, \dots, M_m\}$, and local dataset sizes $\{|D_1|, \dots, |D_m|\}$ of $m$ clients $C$; filtering threshold $\beta$, clustering sensitivity $\gamma$. }
    \KwOutput{Aggregated model $W_G$.}
    $C_\textsc{benign} \leftarrow poisoning\_filtering(C, W, M, \beta, \gamma)$ \;
    \ForEach{parameter pack $P_i$ in global model}{
        $S_i \leftarrow \emptyset$ \;
        \ForEach{client $j$ in $C_\textsc{benign}$}{
            \If{$M_j(P_i) = 1$}{
                Add client $j$ to $S_i$ \;
            }
        }
        $A_{P_i} = \sum_{j \in S_i} \frac{|D_j|}{\sum_{k=1}^{m}|D_k|}$ \;
        Initialize aggregated model parameter $w_i^{agg} = 0$ \;
        \ForEach{client $j \in S_i$}{
            $W_j(P_i) \gets W_j(P_i) * \frac{|D_j|}{\sum_{k=1}^{m}|D_k|}$\;
            Normalize $W_j(P_i) \gets W_j(P_i) / A_{P_i}$\;
            $w_i^{agg} += W_j(P_i)$ \;
        }
         Set $W_G(P_i) = w_i^{agg}$\;
    }
    \Return $W_G$ 

\end{algorithm}

\section{Proof of Theorem 1 (Vulnerability Analysis)}
\label{sec:proof_thm1}

In this section, we provide the formal proof for Theorem 1, which quantifies the effectiveness of poisoning attacks within the sparsified aggregation framework. Our analysis substantiates the claim that sparsification introduces a unique vulnerability by allowing attackers to concentrate their influence on specific parameter subsets.

\subsection{Definition and Metrics}
As stated in Theorem 1, we define the attack effectiveness $\rho$ as the squared Euclidean distance between the poisoned global model $W_G$ and the ideal benign model $W'_G$:
\begin{equation}
\rho = \|W_G - W'_G\|_2^2
\end{equation}
This metric serves as a proxy for the total deviation induced by malicious participants.

\subsection{Bounding the Deviation}

To analyze the attack effectiveness, we first formalize the sparsified aggregation process from a mathematical perspective. The aggregation is performed on a pack-by-pack basis. Let's consider a single parameter pack $p$. The set of clients contributing to this pack is $S_p = \{i \in C \mid M_i(p) = 1\}$, where $M_i(p)$ is the sparse index mask for client $i$ at pack $p$. The size of this set is $N_p = |S_p|$.

This set of contributors can be split into benign and malicious subsets: $S_p^B = S_p \cap C_B$ and $S_p^A = S_p \cap C_A$. Let their sizes be $N_p^B = |S_p^B|$ and $N_p^A = |S_p^A|$, respectively, with $N_p = N_p^B + N_p^A$.

For simplicity, we model the aggregation for pack $p$ (lines 10-14 in Algorithm~\ref{alg:sa}) as a weighted average over the contributing clients. The poisoned global model parameter for pack $p$ is:
\begin{equation}
W_G(p) = \frac{1}{N_p} \left( \sum_{j \in S_p^B} W_j(p) + \sum_{a \in S_p^A} W_a(p) \right)
\end{equation}
where $W_j(p)$ and $W_a(p)$ are the parameters for pack $p$ from a benign client $j$ and a malicious client $a$, respectively. For this analysis, we assume uniform weights for simplicity, though the principle holds for the weighted scheme in Algorithm~\ref{alg:sa}.

The ideal benign model for the same pack $p$ would be aggregated only over the benign contributors:
\begin{equation}
W'_G(p) = \frac{1}{N_p^B} \sum_{j \in S_p^B} W_j(p)
\end{equation}

The core of the attack lies in the malicious parameters $W_a(p)$. We can model a malicious update as a deviation from the ideal benign update. We assume the attacker's update for pack $p$ is bounded in its deviation from the ideal benign average for that pack. Let $\epsilon$ be the expected magnitude of perturbation for any parameter pack an attacker chooses to poison. This is formulated as:
\begin{equation}
\mathbb{E}[\|W_a(p) - W'_G(p)\|_2] \le \epsilon
\label{eq:perturbation_bound}
\end{equation}
for any attacker $a \in S_p^A$. This value $\epsilon$ represents the potency of the malicious craft. A larger $\epsilon$ signifies a more aggressive attack.

\subsection{Proof of the Attack Effectiveness}

With the definitions above, we can now derive an upper bound for the attack effectiveness $\rho$. We start by analyzing the deviation for a single pack $p$.
% \begin{equation}
% \begin{split}
% W_G(p) - W'_G(p) &= \frac{1}{N_p} \left( \sum_{j \in S_p^B} W_j(p) + \sum_{a \in S_p^A} W_a(p) \right) - W'_G(p) \\
% &= \frac{1}{N_p} \left( N_p^B W'_G(p) + \sum_{a \in S_p^A} W_a(p) \right) - W'_G(p) \\
% &= \left(\frac{N_p^B}{N_p} - 1\right) W'_G(p) + \frac{1}{N_p} \sum_{a \in S_p^A} W_a(p) \\
% &= \frac{-N_p^A}{N_p} W'_G(p) + \frac{1}{N_p} \sum_{a \in S_p^A} W_a(p) \\
% &= \frac{1}{N_p} \sum_{a \in S_p^A} (W_a(p) - W'_G(p))
% \end{split}
% \label{eq:pack_deviation}
% \end{equation}
\begin{equation}
\begin{split}
W_G(p) - W'_G(p) 
&= \frac{1}{N_p} \left( \sum_{j \in S_p^B} W_j(p) + \sum_{a \in S_p^A} W_a(p) \right) \\
&\qquad - W'_G(p) \\
&= \frac{1}{N_p} \left( N_p^B W'_G(p) + \sum_{a \in S_p^A} W_a(p) \right) \\
&\qquad - W'_G(p) \\
&= \left(\frac{N_p^B}{N_p} - 1\right) W'_G(p) + \frac{1}{N_p} \sum_{a \in S_p^A} W_a(p) \\
&= \frac{-N_p^A}{N_p} W'_G(p) + \frac{1}{N_p} \sum_{a \in S_p^A} W_a(p) \\
&= \frac{1}{N_p} \sum_{a \in S_p^A} (W_a(p) - W'_G(p))
\end{split}
\label{eq:pack_deviation}
\end{equation}

This equation elegantly captures the error for a single pack: it is the average deviation of the malicious contributors from the ideal benign model, scaled by the malicious-to-total contributor ratio for that pack.

Now, we bound the squared norm of this deviation using the triangle inequality and our perturbation bound from Equation~\ref{eq:perturbation_bound}:
% \begin{equation}
% \begin{split}
% \|W_G(p) - W'_G(p)\|_2^2 &= \left\| \frac{1}{N_p} \sum_{a \in S_p^A} (W_a(p) - W'_G(p)) \right\|_2^2 \\
% &\le \frac{1}{N_p^2} \left( \sum_{a \in S_p^A} \|W_a(p) - W'_G(p)\|_2 \right)^2 \\
% &\le \frac{1}{N_p^2} (N_p^A \epsilon)^2 = \left(\frac{N_p^A}{N_p}\right)^2 \epsilon^2
% \end{split}
% \end{equation}
\begin{equation}
\begin{split}
\|W_G(p) - W'_G(p)\|_2^2 &= \left\| \frac{1}{N_p} \sum_{a \in S_p^A} (W_a(p) - W'_G(p)) \right\|_2^2 \\
&\le \frac{1}{N_p^2} \left( \sum_{a \in S_p^A} \|W_a(p) - W'_G(p)\|_2 \right)^2 \\
&\le \frac{1}{N_p^2} (N_p^A \epsilon)^2 = \left(\frac{N_p^A}{N_p}\right)^2 \epsilon^2
\end{split}
\end{equation}

The total attack effectiveness $\rho$ is the sum of these squared deviations over all parameter packs:
\begin{equation}
\rho = \sum_p \|W_G(p) - W'_G(p)\|_2^2 \le \sum_p \left(\frac{N_p^A}{N_p}\right)^2 \epsilon^2
\label{eq:bound}
\end{equation}

Let us define the \textit{malicious contributor ratio} for pack $p$ as $f_p = N_p^A / N_p$. This ratio represents the fraction of clients that contributed to pack $p$ who were malicious. The final bound on the attack effectiveness is:
\begin{equation}
\rho \le \epsilon^2 \sum_p f_p^2
\label{eq:final_bound}
\end{equation}

\noindent\textbf{Implications.} This result highlights a critical vulnerability unique to communication-efficient FL using sparsification. In standard FL, the overall attacker ratio $|C_A|/|C|$ dilutes the attack's impact across all model parameters relatively uniformly. However, Equation~\ref{eq:final_bound} shows that in sparsified FL, the attack's impact is determined by the malicious contributor ratio $f_p$ \textit{at each parameter pack}. Adversaries can exploit this by coordinating their sparse index masks ($M_a$) to all contribute to a small, critical subset of parameter packs. For these targeted packs, they can drive $f_p$ close to 1, thus exerting immense influence and causing significant model deviation, even if the overall attacker ratio $|C_A|/|C|$ is low. This demonstrates that sparsification, while beneficial for efficiency, introduces a new attack vector that must be addressed by specialized defense mechanisms like SafeSparse.

\section{Proof of Theorem 2 (Convergence Analysis)}
\label{sec:proof_thm2}

In this section, we provide the detailed derivation for the convergence guarantee of SafeSparse presented in Theorem 2. We analyze how the robust aggregation mechanism interacts with sparsification noise and adversarial bias to ensure model stability.

\subsection{Assumptions and Preliminaries}
To facilitate the analysis, we state the following standard assumptions used in Federated Learning literature:

\begin{assumption}[L-smoothness]
The local cost functions $F_i(w)$ are all $L$-smooth: $\|\nabla F_i(w) - \nabla F_i(v)\| \le L\|w-v\|$ for any $w, v$.
\end{assumption}

\begin{assumption}[Bounded Variance and Heterogeneity]
The local stochastic gradients have a variance bounded by $\sigma^2$. Furthermore, the data heterogeneity (Non-IIDness) is bounded by $G^2$, such that $\mathbb{E}\|\nabla F_i(w) - \nabla f(w)\|^2 \le G^2$.
\end{assumption}

\begin{assumption}[Top-$k$ Compression Property]
The sparsification operator $S(\cdot)$ satisfies the compression property $\|S(x) - x\|^2 \le (1-\alpha)\|x\|^2$, where $\alpha = k/d$ is the fraction of parameters retained.
\end{assumption}

\begin{assumption}[Filtering Integrity]
Let $\mathcal{C}_B$ be the set of clients selected by SafeSparse. There exists a filtering error $\delta(\beta, \gamma)$ such that the residual bias from missed malicious updates or excluded benign updates is bounded: $\|\mathbb{E}[\Delta w_{agg}] - \Delta w_{ideal}\| \le \delta$, where $\delta$ is monotonically decreasing with the accuracy of the Jaccard and Sign-based inspection.
\end{assumption}

\subsection{Restatement of Theorem 2}

Under Assumptions 1-4, let $\eta_t = \frac{8}{\mu(t+a)}$ be the learning rate for a $\mu$-strongly convex objective. The global model $W_G^T$ after $T$ rounds of SafeSparse satisfies:
\begin{equation}
\mathbb{E}[\|W_G^T - W^*\|^2] \le \frac{D}{T+a} + \underbrace{\mathcal{O}\left(\frac{(1-\alpha)G^2}{\alpha^2}\right)}_{\text{Sparsification Error}} + \underbrace{\mathcal{O}\left(\frac{\rho}{\mu^2}\right)}_{\text{Residual Attack Impact}}
\end{equation}
where $D$ is a constant related to the initial distance, $\alpha$ is the sparsity ratio, and $\rho$ is the attack effectiveness defined in Equation~\ref{eq:bound}.

\subsection{Formal Derivation}

To prove the convergence of SafeSparse, we begin by analyzing the evolution of the expected distance between the global model and the optimal solution across iterations. Given the update rule $W_{t+1} = W_t - \eta_t \mathcal{G}_t$, where $\mathcal{G}_t$ represents the robustly aggregated sparse gradient, we expand the squared Euclidean distance to the optimum $W^*$ as follows:
\begin{equation}
\|W_{t+1} - W^*\|^2 = \|W_t - W^*\|^2 - 2\eta_t \langle \mathcal{G}_t, W_t - W^* \rangle + \eta_t^2 \|\mathcal{G}_t\|^2
\label{eq:expansion}
\end{equation}

The core of the analysis lies in bounding the inner product term, which reflects the alignment of the aggregated update with the descent direction. In the presence of poisoning attacks, the aggregated gradient $\mathcal{G}_t$ deviates from the true population gradient $\nabla f(W_t)$. By substituting the attack effectiveness bound derived in Section~\ref{sec:proof_thm1}, we can quantify this deviation as $\|\mathbb{E}[\mathcal{G}_t] - \nabla f(W_t)\|^2 \le \mathcal{O}(\rho)$. SafeSparse's dual-inspection mechanism—specifically the Jaccard similarity and Sign clustering—is instrumental here; it minimizes $\rho$ by effectively reducing the malicious contributor ratio $f_p$ across all parameter packs, thereby constraining the adversarial bias within the inner product.

Simultaneously, we must account for the variance introduced by the sparsification process. While the Top-$k$ operator inherently introduces a residual error proportional to $(1-\alpha)$, SafeSparse ensures that only updates with high structural consensus (verified via Jaccard similarity) are aggregated. This filtering prevents the divergence of the term $\eta_t^2 \|\mathcal{G}_t\|^2$ by ensuring that the sparsity masks of contributing clients remain coherent, even under statistical heterogeneity. 

Finally, by combining these bounds into the recursion in Equation~\ref{eq:expansion} and applying a decaying learning rate $\eta_t = \frac{8}{\mu(t+a)}$, the inequality satisfies the conditions for a standard inductive proof for strongly convex functions. This leads to the conclusion that $W_G^T$ converges to a fixed error ball at a rate of $\mathcal{O}(1/T)$. The radius of this error ball is explicitly determined by the sparsity ratio $\alpha$ and the residual poisoning impact $\rho$ that bypasses the filters, demonstrating that SafeSparse achieves a stable consensus in adversarial sparse environments.

\section{Evaluations}

\begin{figure*}[htb]
\centering
\includegraphics[width=0.9\textwidth,trim=0 5pt 0 5pt,clip]{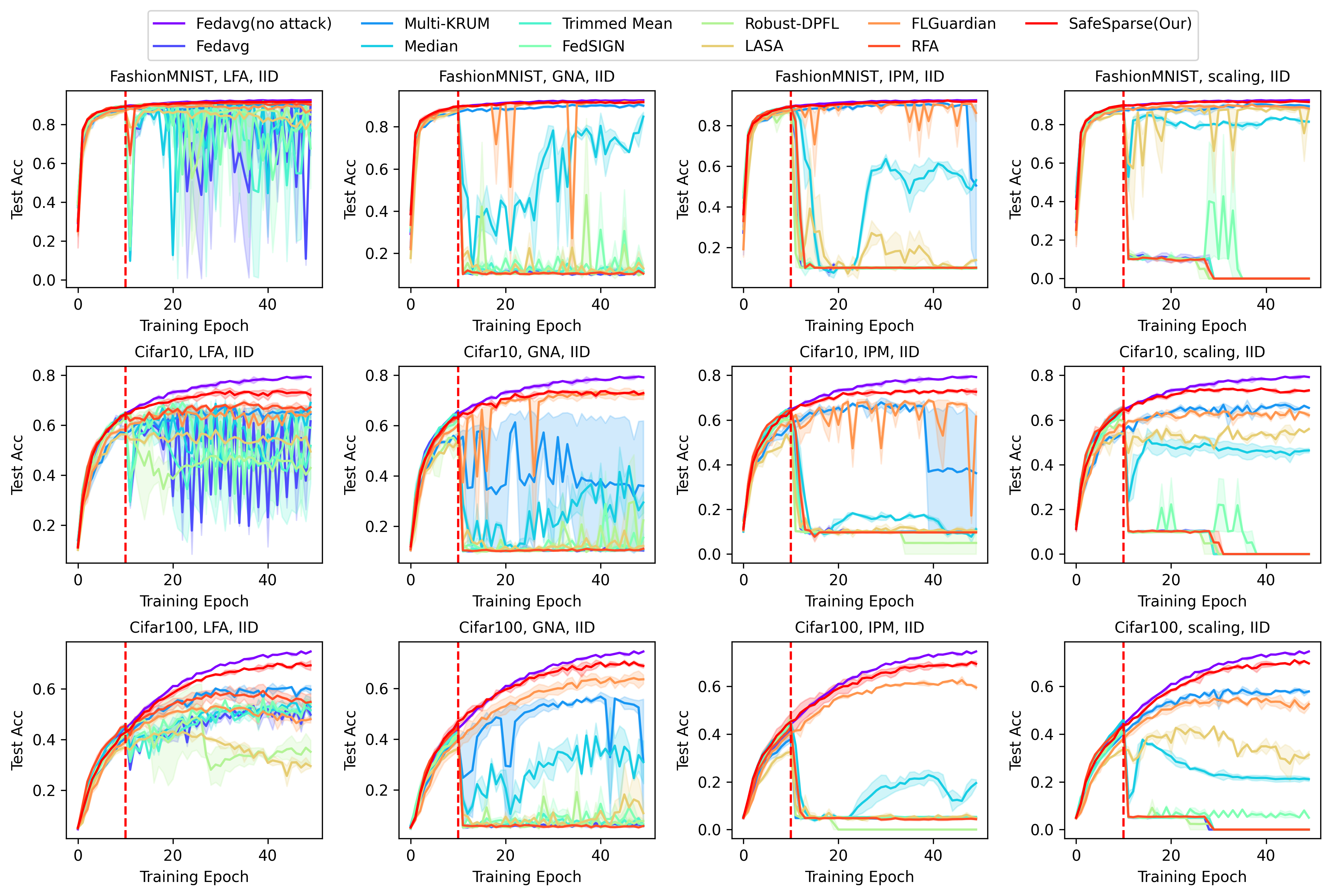}
% \vspace{-3mm}
\caption{Comparison of defense effectiveness across various approaches, evaluated on FashionMNIST, CIFAR-10, and CIFAR-100 under Label Flip Attack (LFA), Gaussian Noise Attack (GNA), Inner Product Manipulation (IPM), and Scaling Attack in an \textbf{IID SETTING}. The default attacker ratio is fixed at 0.4, and the top-$k$ sparsification rate is set to 0.5. Top-1 accuracy is used as the evaluation metric for FashionMNIST and CIFAR-10, while top-5 accuracy is adopted for CIFAR-100.}
\label{iid}
% \vspace{-5mm}
\end{figure*}

\begin{table*}[h]
\centering
\caption{Comparison of defense effectiveness across different $\beta$ settings (IID Distribution).}
\label{table:beta_iid}
\small
\begin{tabular}{llccccc}
\toprule
\textbf{Dataset} & \textbf{Attack} & \textbf{$\beta=0.20$} & \textbf{$\beta=0.40$} & \textbf{$\beta=0.60$} & \textbf{$\beta=0.80$} & \textbf{$\beta=1.00$} \\
\midrule
\multirow{4}{*}{FashionMNIST} 
 & LFA     & 0.9164 & \textbf{0.9175} & \textit{0.9132} & 0.9156 & 0.9148 \\
 & GNA     & \textbf{0.9218} & \textit{0.9102} & 0.9113 & 0.9194 & 0.9166 \\
 & Scaling & 0.9155 & \textbf{0.9201} & 0.9127 & 0.9165 & \textit{0.9126} \\
 & IPM     & \textbf{0.9177} & 0.9146 & 0.9141 & 0.9140 & \textit{0.8712} \\
\midrule
\multirow{4}{*}{CIFAR-10} 
 & LFA     & \textbf{0.7378} & 0.7279 & 0.7319 & 0.7324 & \textit{0.7246} \\
 & GNA     & 0.7293 & 0.7492 & \textbf{0.7511} & \textit{0.7140} & 0.7288 \\
 & Scaling & 0.7420 & 0.7368 & 0.7399 & \textbf{0.7446} & \textit{0.7022} \\
 & IPM     & 0.7209 & 0.7049 & \textbf{0.7416} & 0.6461 & \textit{0.6371} \\
\midrule
\multirow{4}{*}{CIFAR-100} 
 & LFA     & 0.6814 & 0.6876 & 0.6887 & \textbf{0.6897} & \textit{0.6262} \\
 & GNA     & 0.6864 & 0.6882 & 0.6981 & \textbf{0.7008} & \textit{0.6904} \\
 & Scaling & 0.6789 & \textbf{0.7006} & 0.6906 & 0.6976 & \textit{0.6425} \\
 & IPM     & 0.5748 & 0.7014 & \textbf{0.7096} & 0.6926 & \textit{0.5736} \\
\bottomrule
\end{tabular}
\end{table*}

\begin{table*}[t]
\centering
\caption{Comparison of defense effectiveness across different $\gamma$ settings (IID Distribution).}
\label{table:gamma_iid}
\small
\begin{tabular}{llccccccc >{\columncolor{gray!15}}c >{\columncolor{gray!15}}c}
\toprule
\textbf{Dataset} & \textbf{Attack} & \textbf{0.10} & \textbf{0.15} & \textbf{0.20} & \textbf{0.25} & \textbf{0.30} & \textbf{0.35} & \textbf{0.40} & \textbf{0.45} & \textbf{0.50} \\
\midrule
\multirow{4}{*}{F-MNIST} 
 & LFA     & 0.9181 & 0.9196 & \textbf{0.9208} & 0.9156 & 0.9163 & 0.9128 & 0.9185 & 0.6692 & 0.7368 \\
 & GNA     & 0.9133 & 0.9171 & 0.9179 & 0.9156 & 0.9167 & \textbf{0.9213} & 0.9194 & 0.1038 & 0.1019 \\
 & Scaling & 0.9160 & 0.9133 & 0.9185 & 0.9109 & \textbf{0.9197} & 0.9108 & 0.9146 & 0.1039 & 0.1042 \\
 & IPM     & 0.7926 & 0.9144 & 0.9122 & 0.9137 & \textbf{0.9166} & 0.8959 & 0.9139 & 0.1017 & 0.1028 \\
\midrule
\multirow{4}{*}{CIFAR-10} 
 & LFA     & 0.7271 & 0.7297 & 0.7227 & 0.6602 & 0.7479 & \textbf{0.7574} & 0.7492 & 0.4215 & 0.5004 \\
 & GNA     & \textbf{0.7429} & 0.7143 & 0.7340 & 0.7388 & 0.7377 & 0.7278 & 0.7237 & 0.1029 & 0.1071 \\
 & Scaling & \textbf{0.7442} & 0.7370 & 0.7328 & 0.7375 & 0.7146 & 0.7301 & 0.7273 & 0.1013 & 0.0973 \\
 & IPM     & 0.4796 & 0.7148 & 0.7060 & 0.6479 & 0.7160 & 0.7198 & \textbf{0.7208} & 0.1031 & 0.1031 \\
\midrule
\multirow{4}{*}{CIFAR-100} 
 & LFA     & 0.6969 & 0.6943 & 0.6878 & 0.6798 & 0.6795 & 0.7055 & \textbf{0.7126} & 0.4994 & 0.4923 \\
 & GNA     & 0.6891 & 0.6929 & 0.6898 & \textbf{0.7024} & 0.6853 & 0.6812 & 0.6849 & 0.0594 & 0.0634 \\
 & Scaling & 0.6853 & \textbf{0.7086} & 0.6822 & 0.6746 & 0.6897 & 0.6953 & 0.6856 & 0.0862 & 0.0532 \\
 & IPM     & 0.6474 & 0.6486 & 0.6125 & 0.5579 & \textbf{0.6562} & 0.6201 & 0.6032 & 0.0457 & 0.0460 \\
\bottomrule
\end{tabular}
\end{table*}

\begin{figure}[h]
\centering 
\includegraphics[width=\linewidth,trim=0 5pt 0 5pt,clip]{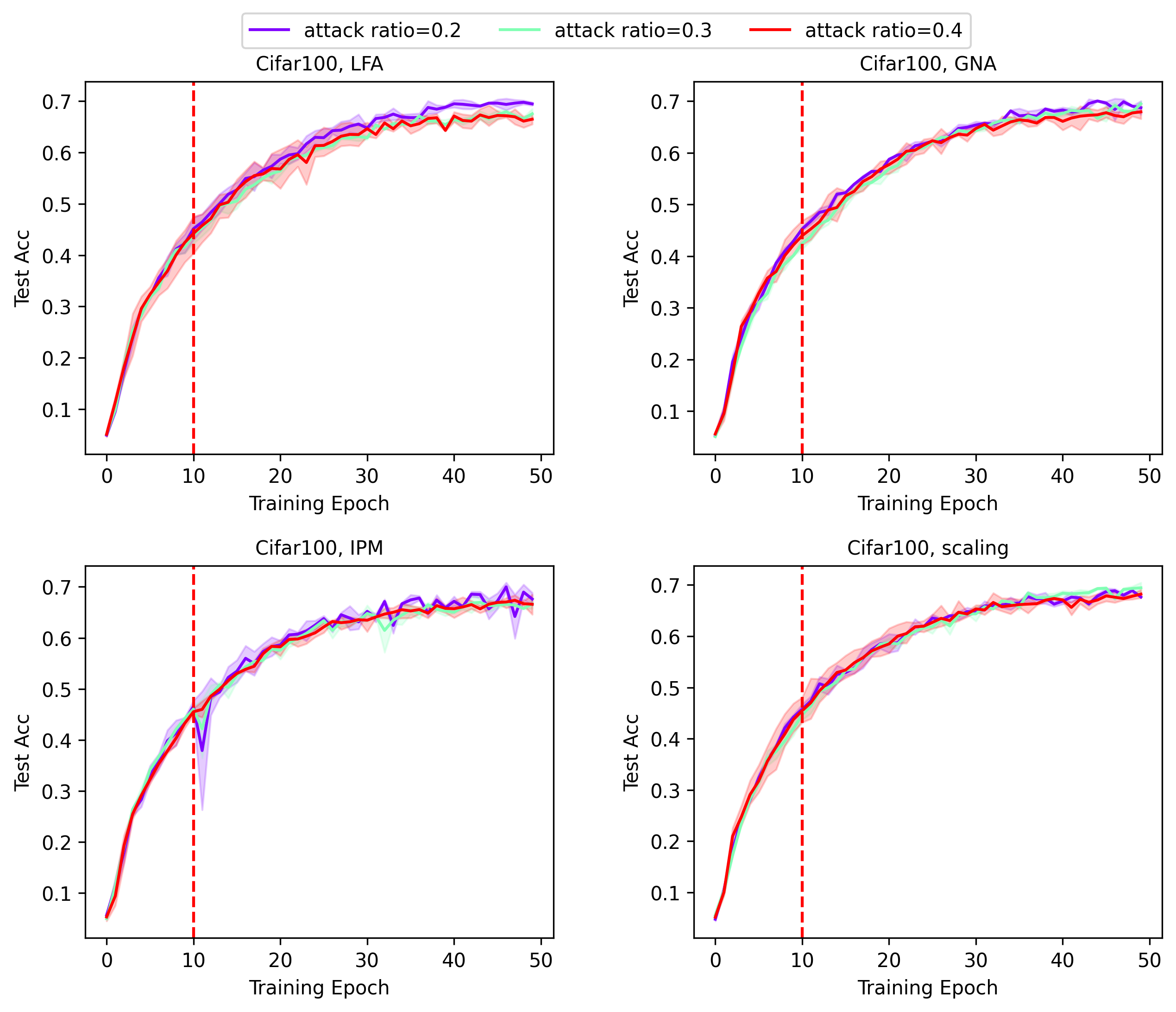}
% \vspace{-8mm}
\caption{Comparison of defense effectiveness across different attack ratios, evaluated on CIFAR-100 under LFA, GNA, IPM, and Scaling attacks. The top-$k$ ratio is set to 0.5.}
\label{attack_ratio}
% \vspace{-3mm}
\end{figure}

\begin{figure}[h]
\centering
\includegraphics[width=\linewidth,trim=0 5pt 0 5pt,clip]{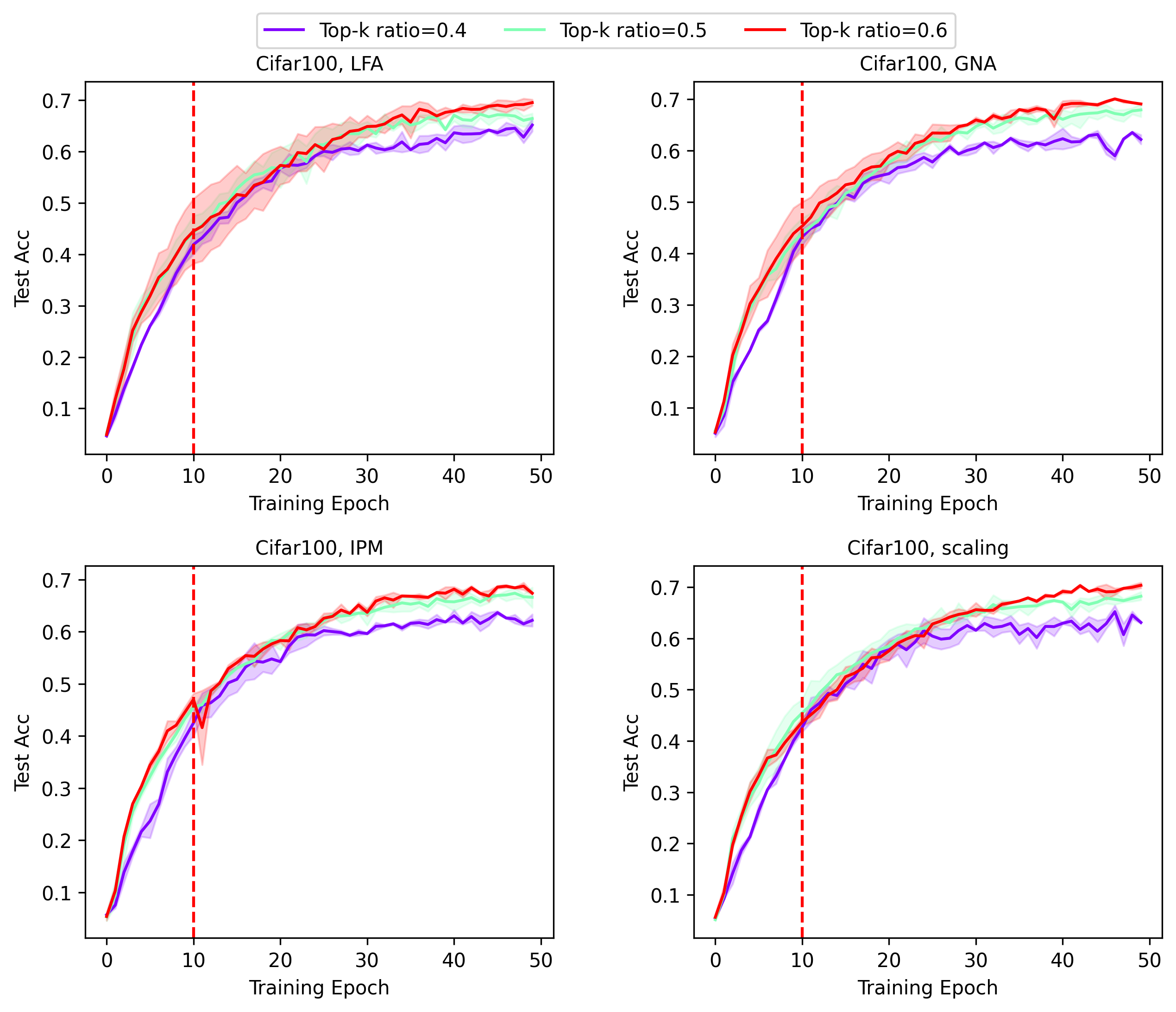}
% \vspace{-8mm}
\caption{Comparison of defense effectiveness across different top-k ratios, evaluated on CIFAR-100 under LFA, GNA, IPM attack, and scaling attack. The attacker ratio is set to 0.4.}
\label{topk}
% \vspace{-5mm}
\end{figure}

\subsection{Baselines}

To provide a comprehensive benchmark for SafeSparse, we compare it against a diverse set of state-of-the-art robust aggregation rules and defense mechanisms. These baselines range from classical statistical robust aggregators to recent advanced methods based on clustering and signature detection. The detailed technical descriptions and their roles in our experimental setup are summarized as follows:

\begin{itemize}
    \item \textbf{FedAVG \cite{mcmahan2017communication}:} 
    A widely used aggregation method that computes a weighted average of model updates from multiple clients to generate a global model. The weights are based on each client's data size.

    \item \textbf{Multi-KRUM \cite{blanchard2017machine}:} Calculates the Euclidean distance between each client update and all others. For each update, the nearest $m-n-1$ distances are summed to compute a $K_r$ score. The $k$ updates with the smallest $K_r$ scores form the set of honest updates ($H$), which is then aggregated using FedAVG.

    \item \textbf{Median \cite{yin2018byzantine}:} Independently calculates the median value for each model update across all client updates and uses it as the global model update.
    
    \item \textbf{Trimmed Mean \cite{yin2018byzantine}:} Sorts client updates and removes the largest and smallest $k\%$. The mean of the remaining values is used as the global model update. In our experiments, $k$ is set to 10 by default.

    \item \textbf{FedSign \cite{guo2023fedsign}:} Identifies malicious clients by measuring cosine similarity between their signature vectors and calculating attack density. Clients with an attack density below average are considered honest, and their updates are aggregated using FedAVG.
    
    \item \textbf{Robust-DPFL \cite{qi2024towards}:} Computes a detection score for each client by averaging its model parameters. Detection scores are clustered into two groups using K-means, and only updates from the cluster with higher average detection scores are aggregated.
    
    \item \textbf{LASA \cite{xu2025achieving}:} An approach that combines pre-aggregation sparsification with layer-wise adaptive aggregation. It first sparsifies updates to reduce the impact of malicious parameters, then uses a layer-wise adaptive filter that leverages both magnitude and direction metrics to select and aggregate benign layers.
    
    \item \textbf{FLGuardian \cite{zhou2025flguardian}:} A layer-space defense method that detects benign clients for each layer using a combination of pairwise cosine/Euclidean distances and a clustering algorithm. It then assigns a trust score to each client and aggregates updates from clients with the highest scores.
    
    \item \textbf{RFA \cite{pillutla2022robust}:} A robust aggregation approach that uses the geometric median of client updates to produce the global model. The geometric median is computed efficiently using a Weiszfeld-type algorithm.
\end{itemize}

In our evaluation, we consider a threat model where a fraction of malicious clients (up to 40\%) attempt to compromise the global model's integrity or prevent its convergence. We implement four representative poisoning attacks that target different stages of the local training and update process. The specific implementations are detailed below:

\begin{itemize}  
    \item \textbf{LFA:} The attacker changes the original label $l$ of a training sample to $M-l-1$, where $M$ is the total number of labels (i.e., categories) in the dataset. The attacker then performs local training on this modified dataset.  
    \item \textbf{GNA:} Using the mean and variance of local model, the attacker generates a Gaussian noise model matching the size of the local model and uploads it to the server.  
    \item \textbf{IPM:} The attacker manipulates the inner product between true gradients and robust aggregated gradients to be negative, disrupting global model convergence.  
    \item \textbf{Scaling:} The attacker amplifies local model updates by a large factor before uploading them to the server.  
\end{itemize}

\subsection{Model Performance and Defense Effectiveness under IID Distribution}

To further validate the robustness of SafeSparse, we evaluate its performance under the IID (Independent and Identically Distributed) setting across three benchmark datasets: FashionMNIST, CIFAR-10, and CIFAR-100. As illustrated in Figure \ref{iid}, SafeSparse consistently achieves superior test accuracy compared to state-of-the-art defense mechanisms under four distinct types of poisoning attacks.

\subsection{Ablation Studies under IID Distribution}
This section supplements the main text with ablation results under the IID distribution, as shown in Table~\ref{table:beta_iid} and Table~\ref{table:gamma_iid}. Overall, the trends are consistent with the Non-IID analysis but exhibit lower variance due to the higher intrinsic similarity among benign updates.

For the threshold $\beta$, performance remains robust until $\beta=1.0$, where a slight decline occurs as some beneficial gradient information is discarded. Regarding $\gamma$, the defense is highly effective when $\gamma \le 0.40$, but accuracy collapses to near-random levels ($\approx 10\%$) beyond this threshold. This confirms that the optimal hyperparameter range is relatively distribution-agnostic, which simplifies parameter tuning in practical federated learning deployments.

\subsection{Impact of Attacker Ratio}

To further investigate the impact of the attack ratio on SafeSparse, we evaluated the defense effectiveness of various attack strategies under different attack ratios (0.2, 0.3, and 0.4) on the CIFAR-100 dataset. It is important to note that, in our Threat Assumption, the attacker ratio will not exceed 50\%. As shown in Figure~\ref{attack_ratio}, the results indicate that the defense methods are effective across all attack ratios, with no significant changes in performance. The test accuracy remains relatively stable regardless of the increase in attack ratio, demonstrating that our proposed defenses can effectively resist model poisoning attacks at various levels.

\subsection{Impact of Top-k Sparsification Ratio}

To investigate how the top-$k$ sparsification ratio affects SafeSparse under various attack strategies, we conducted experiments at different sparsification levels on the CIFAR-100 dataset. Specifically, the top-$k$ sparsification ratio was set between 0.4 and 0.6 in our experiments. As shown in Figure~\ref{topk}, the final convergence results indicate that the attacker has no significant impact on our proposed SafeSparse across all sparsification ratios. SafeSparse demonstrates strong resilience against four types of model poisoning attacks. However, lower sparsification ratios may lead to unstable attacker detection.

% \section*{Ethical Statement}

% There are no ethical issues.

% \section*{Acknowledgments}

%% The file named.bst is a bibliography style file for BibTeX 0.99c

\end{document}